\documentstyle[galley,epsf]{mn}

\newif\ifAMStwofonts

\ifoldfss
  \ifCUPmtlplainloaded \else
    \NewTextAlphabet{textbfit} {cmbxti10} {}
    \NewTextAlphabet{textbfss} {cmssbx10} {}
    \NewMathAlphabet{mathbfit} {cmbxti10} {} 
    \NewMathAlphabet{mathbfss} {cmssbx10} {} 
  \fi
  \ifAMStwofonts
    \ifCUPmtlplainloaded \else
      \NewSymbolFont{upmath} {eurm10}
      \NewSymbolFont{AMSa} {msam10}
      \NewMathSymbol{\upi}     {0}{upmath}{19}
      \NewMathSymbol{\umu}     {0}{upmath}{16}
      \NewMathSymbol{\upartial}{0}{upmath}{40}
      \NewMathSymbol{\leqslant}{3}{AMSa}{36}
      \NewMathSymbol{\geqslant}{3}{AMSa}{3E}

      \let\leq=\leqslant 
      \let\geq=\geqslant 
    \fi
  \fi
\fi 

\ifnfssone
  \newmathalphabet{\mathit}
  \addtoversion{normal}{\mathit}{cmr}{m}{it}
  \addtoversion{bold}{\mathit}{cmr}{bx}{it}
  \newmathalphabet{\mathbfit} 
  \addtoversion{normal}{\mathbfit}{cmr}{bx}{it}
  \addtoversion{bold}{\mathbfit}{cmr}{bx}{it}
  \newmathalphabet{\mathbfss} 
  \addtoversion{normal}{\mathbfss}{cmss}{bx}{n}
  \addtoversion{bold}{\mathbfss}{cmss}{bx}{n}
  \ifAMStwofonts
    \ifCUPmtlplainloaded \else
      \UseAMStwoboldmath
      \makeatletter
      \new@mathgroup\upmath@group
      \define@mathgroup\mv@normal\upmath@group{eur}{m}{n}
      \define@mathgroup\mv@bold\upmath@group{eur}{b}{n}
      \edef\UPM{\hexnumber\upmath@group}
      \new@mathgroup\amsa@group
      \define@mathgroup\mv@normal\amsa@group{msa}{m}{n}
      \define@mathgroup\mv@bold\amsa@group{msa}{m}{n}
      \edef\AMSa{\hexnumber\amsa@group}
      \makeatother
      \mathchardef\upi="0\UPM19
      \mathchardef\umu="0\UPM16
      \mathchardef\upartial="0\UPM40
      \mathchardef\leqslant="3\AMSa36
      \mathchardef\geqslant="3\AMSa3E

      \let\leq=\leqslant 
      \let\geq=\geqslant 
    \fi
  \fi
\fi 

\ifnfsstwo
  \DeclareMathAlphabet{\mathbfit}{OT1}{cmr}{bx}{it}
  \SetMathAlphabet\mathbfit{bold}{OT1}{cmr}{bx}{it}
  \DeclareMathAlphabet{\mathbfss}{OT1}{cmss}{bx}{n}
  \SetMathAlphabet\mathbfss{bold}{OT1}{cmss}{bx}{n}
  \ifAMStwofonts
    \ifCUPmtlplainloaded \else
      \DeclareSymbolFont{UPM}{U}{eur}{m}{n}
      \SetSymbolFont{UPM}{bold}{U}{eur}{b}{n}
      \DeclareSymbolFont{AMSa}{U}{msa}{m}{n}
      \DeclareMathSymbol{\upi}{0}{UPM}{"19}
      \DeclareMathSymbol{\umu}{0}{UPM}{"16}
      \DeclareMathSymbol{\upartial}{0}{UPM}{"40}
      \DeclareMathSymbol{\leqslant}{3}{AMSa}{"36}
      \DeclareMathSymbol{\geqslant}{3}{AMSa}{"3E}

      \let\leq=\leqslant 
      \let\geq=\geqslant 
    \fi
  \fi
\fi 

\ifCUPmtlplainloaded \else
  \ifAMStwofonts \else 
    \def\upi{\pi}
    \def\umu{\mu}
    \def\upartial{\partial}
  \fi
\fi

\title{The 1995-1996 Decline of R Coronae Borealis -- High Resolution Optical
Spectroscopy}
\author[N. Kameswara Rao et al.]
       {N. Kameswara Rao,$^1$ David L. Lambert,$^2$ Mark T. Adams,$^3$ 
        David R. Doss,$^3$
\newauthor
       Guillermo Gonzalez,$^4$ Artie P. Hatzes,$^2$ C. Ren\'{e}e James,$^2$
       C. M. Johns-Krull,$^5$ 
\newauthor
       R. Earle Luck,$^6$ Gajendra Pandey,$^1$ Klaus Reinsch,$^7$
       Jocelyn Tomkin,$^2$
\newauthor
        and Vincent M. Woolf$^2$\\
       $^1$Indian Institute of Astrophysics, Bangalore 560034, India\\
       $^2$Department of Astronomy, University of Texas, Austin, TX 78712-1083, USA\\
        $^3$McDonald Observatory, Fort Davis, TX 79734-1337, USA\\
        $^4$Department of Astronomy, University of Washington,
 P.O. Box 351589, Seattle, WA 98195-1580, USA\\
         $^5$Space Science Laboratories, University of California,
         Berkeley, CA 94720-7450, USA\\
        $^6$Department of Astronomy, Case Western Reserve University, Cleveland,
        OH 44106-7215\\
        $^7$Universit\"{a}ts-Sternwarte, Georg-August Universit\"{a}t, G\"{o}ttingen,  37083 Germany\\}
    
\date{Accepted .
      Received ;
      in original form 1999 }

\pagerange{\pageref{firstpage}--\pageref{lastpage}}
\pubyear{1999}

\begin{document}

\maketitle

\label{firstpage}

\begin{abstract}

 A set of high-resolution optical spectra of R\,CrB acquired before,
 during, and after
its 1995-1996 decline is discussed. All of the components reported
from earlier declines are seen. This novel dataset provides new
information on these components including several aspects not
previously seen in declines of R\,CrB and other RCBs. 
In the latter category is the discovery that the decline's onset 
 is marked by distortions of absorption lines
of high-excitation lines, and quickly followed by emission in these
and in low excitation
lines. This `photospheric trigger' implies that dust causing the
decline is formed close to the star.  These emission lines fade quickly.
 After 1995 November 2, low excitation narrow (FWHM $\sim 12$ km s$^{-1}$)
emission lines remain. These appear to be a permanent feature, slightly
blue-shifted from the systemic velocity, and 
unaffected by the decline except for a late and slight decrease of 
flux at minimum light. The location of the  warm dense gas providing
these lines is uncertain. Absorption lines unaffected by overlying sharp
emission are greatly broadened, weakened, and red-shifted at the faintest
magnitudes when scattered light from the star is a greater contributor than
direct light transmitted through the fresh soot cloud.
  A few broad lines (FWHM $\simeq 300 $ km s$^{-1}$)
are seen at and near minimum light with approximately constant
flux: prominent among these are the
He\,{\sc i} triplet series, Na\,{\sc i} D, and [N\,{\sc ii}] lines. These lines
are blue-shifted by about 30 km s$^{-1}$ relative to the systemic 
velocity with no change in velocity over the several months for which the
lines were seen. It is suggested that these lines, especially the He\,{\sc i}
lines, arise from an accretion
disk around an unseen compact companion, which may be  a low-mass
white dwarf. If so, R\,CrB is similar to the unusual post-AGB star 89\,Her.

\end{abstract}


\section{Introduction}

R Coronae Borealis is the prototype of a class of very rare and peculiar 
supergiant stars
with two distinctive primary traits, one photometric and the
other spectroscopic. Photometrically, an RCB is distinct because it
 declines at unpredictable
times by one to several
magnitudes as a cloud of carbon soot obscures the stellar photosphere
for weeks to months.
Spectroscopically, the distinctive signature of an RCB  is weak Balmer lines
that indicate an atmosphere deficient in hydrogen.
Two fundamental questions about RCBs remain
unanswered: By what evolutionary paths are some stars with their normal
H-rich atmospheres converted to RCBs with  He-rich atmospheres?
What are the physical processes that trigger and control  development
of the unpredictable minima?

In this paper, we discuss spectroscopic observations of
the recent deep and prolonged  minimum of  R CrB and aim to address the
second question.
The decline that seems to have begun on or around 1995 October 2 proved to be
the deepest and longest decline of recent years. Recovery from   
the minimum of the decline  was
slow; even about 1 year after the onset of the decline the star
was  1 magnitude below its normal maximum.
Throughout this period, we were able to obtain 
high-resolution optical spectra of the star at quasi-regular intervals.


Our discussion of this novel dataset provides new insights into the
widely accepted model of RCB declines in which a cloud of carbon soot
obscures the star (O'Keefe 1939). There is convincing empirical
evidence that the cloud is a localized event and not a spherically
symmetric phenomenon. In particular, the infrared excess of the 
star is largely unchanged during the decline showing that a large
dust cloud exists independently of the new decline and that the
largely unobscured star heats this cloud (Feast 1979, 1996).
The dusty cloud with a radius of  100$R_*$ is heated by absorbing
about 10\% to 20\% of the stellar radiation
 (Forrest, Gillett \& Stein 1971,1972;
Rao \& Nandy 1986).

 How and where the soot
condenses has been debated.  If dust is to form under equilibrium conditions
in a quasi-hydrostatic extension of the stellar atmosphere, the
required low temperatures are found only far from the star - say, at 10-20
stellar radii. If regions of the atmosphere are compressed by a shock,
the necessary low temperatures can be  found for a time much closer to the
star - say at 1-2 stellar radii. Observational and theoretical arguments
for the `far' and `near' sites of dust formation are
reviewed by Clayton (1996, also Fadeyev 1986). The initial fading of the
star has been plausibly interpreted as due to the lateral
growth of a dust cloud.

Earlier accounts  of R\,CrB,  RY\,Sgr and V854\,Cen  have
identified the following principal spectroscopic components that
are revealed as an   RCB is obscured by soot (see Clayton
1996 for a general review of  observational characteristics of 
RCBs in and out of decline, and of their evolutionary origins):

\begin{itemize}

\item
{\bf Sharp emission lines.} 
 These lines  appear shortly
after the onset of a decline and disappear just before the return
to maximum light.  Alexander et al. (1972) in an extensive study of
photographic spectra of RY\,Sgr divided these emission lines into
two types - E1 and E2. Payne-Gaposchkin (1963) earlier noted the
two types. A defining characteristic of E1 lines is that they fade
away after about two weeks from a decline's onset. Membership in E1 has
been summarized by the remark that ``this spectrum consists of many lines
of neutral and singly ionized metals'' (Clayton 1996). To this must be
added the comment that the excitation of the E1 spectrum of lines
is higher than that of the E2 spectrum. E2 lines are present throughout
the decline and 
 are likely permanent features.   Prominent in the E2 (and E1) spectrum    
are  lines of ions such as
Sc\,{\sc ii}, Ti\,{\sc ii},  Fe\,{\sc ii},
Y \,{\sc ii}, Ba\,{\sc ii},  as well as neutral atoms, particularly
 Fe\,{\sc i}.
 The emission lines
are slightly blue-shifted 
with respect to the star's systemic velocity.

Accounts of these lines
were given by Payne-Gaposchkin (1963)
and Cottrell, Lawson \& Buchhorn (1990) for R CrB, and Alexander et al.
(1972) for RY Sgr -- see also  Rao \&
 Lambert (1993) on
V854~Cen and Goswami et al.
 (1997) on S Aps, both in deep declines. 

\item
{\bf Broad permitted and forbidden emission lines.} Seen in deep declines,
these are
are much broader than the sharp emission
lines. In the case of V854~Cen, for example,
the width (FWHM) of the broad lines was about 300 km s$^{-1}$ but the
sharp lines were unresolved with a  FWHM less than about 20 km s$^{-1}$ 
(Rao \& Lambert 1993). 

 The first report of  forbidden and permitted broad lines in the spectrum
of an RCB in decline was Herbig's (1949, 1968) discovery of
[O\,{\sc ii}] 3727\AA\ and He\,{\sc i} 3889\AA\
in the spectrum of R CrB. Herbig was unable to determine that the 
line widths differed from that of the Sc\,{\sc ii} {\it et al.} lines but
attribution of the lines to the group of broad lines now seems evident.



\item
{\bf Photospheric absorption lines.} In the early phases of a decline, the
sharp emission lines are superimposed on a photospheric spectrum that
appears largely unchanged for those lines that do not go into emission. In
deep declines, the photospheric spectrum changes. A weakening of the
lines  noted by Herbig (1949) was confirmed and discussed by Payne-Gaposchkin
(1963) and Cottrell et al. (1990) for R CrB. The weakening was attributed
to `veiling', a term implying dilution of the photospheric spectrum
by overlying continuous (or line) emission. In an observation of V854~Cen
in a deep decline, the continuous spectrum was devoid of lines (Rao \&
Lambert 1993).

\item
{\bf Shell absorption components.} During the recovery to maximum light
and into full recovery, blue-shifted broad absorption components of the
Na\,{\sc i} D, and Ca\,{\sc ii} H and K lines are seen. A velocity shift of
-150 km s$^{-1}$ seems typical. Reports of these lines were given by
 Payne-Gaposchkin (1963),  Rao (1974), Cottrell et al. (1990), and Lambert, Rao
\& Giridhar (1990)
 for R CrB, Alexander
et al. (1972) and Vanture \& Wallerstein (1996)
 for RY Sgr, and Clayton et al. (1993) and Rao \& Lambert (1993) for
V854\,Cen. 

\end{itemize}

In this study of R\,CrB's 1995-1996 deep decline, we discuss these
spectroscopic components with an emphasis on novel results.
 With our temporal and spectral coverage,
new detailed information is provided on
all of the previously seen principal components. Since the
decline was preceded by a long interval in which R\,CrB was
at or near maximum light, it is most likely that all of the
spectroscopic changes are associated with the decline and none
are residual effects of earlier declines.
 Following a description of the sequence of spectra acquired
from 1995 to 1996, discussion
is arranged chronologically beginning with descriptive
remarks on the
spectrum of the star prior to the decline, continuing with spectroscopic
changes associated with the onset of the decline, and concluding with
remarks on the several components present in the spectrum from
about mid-decline through the extended period of minimum light and into
the recovery phase. These descriptions are followed by interpretative
remarks on a model of R\,CrB including a suggestion that this star
may be a spectroscopic binary.

\section{Observations}
 
Spectra were obtained at the W. J. McDonald Observatory
with either the 2.7m or the 2.1m reflector.
The light curve of R\,CrB through the decline is shown in the lower
panel of Fig. 1 where the sources of the photometry are identified.
 The upper panel gives radial velocity measurements
and the predicted velocity variation due to pulsation (see below).
The measurements serve to indicate the relation between our observations and
the phase of the decline -- see also Table 1.

\begin{table*}
\centering
\begin{minipage} {140mm}
\caption {Catalogue of spectra of R\,CrB.}
 \begin{tabular}{rlcccccr} \hline 
 &   Date  &    JD - 2440000.0& {Mag.}\footnote{Visual magnitude from AAVSO. V (given to second decimal from Fernie (1997) and Efimov (1997).} &  Telescope & {Observer}%
 \footnote{Observers: DD = David Doss, GG = Guillermo Gonzalez, AH = Artie P. Hatzes,
	     CRJ = C. Ren\'{e}e James, CJK = Chris Johns-
	     Krull,  REL = R. Earle Luck, KR = Klaus
	     Reinsch, JT = Jocelyn Tomkin, VW  = Vincent Woolf}
  &  {Comments}%
\footnote{This gives wavelength interval in \AA\ for the 2.1m spectra.} & \\ \hline 

1995&Jan 24 &  9742.01  &  5.9  &  2.1m   &    GG &   5720 - 7225  &\\
    &Feb 20 &  9768.97  &  5.9  &  2.1m   &    AH &   5590 - 7040 &\\
    &Feb 23 &  9771.92  &  5.9  &  2.1m   &    AH &   5990 - 6960 &\\
    &Mar 17 &  9793.96  &  5.9  &  2.1m   &    AH &   5990 - 6960 &\\
    &Mar 19 &  9796.02  &  5.9  &  2.1m   &    AH &   5990 - 6960 &\\
    &Mar 20 &  9797.00  &  5.9  &  2.1m   &    AH &   5990 - 6960 &\\
    &Apr 14 &  9821.80  &  5.9  &  2.1m   &    REL/VW& 5720 - 7225 &\\
    &Apr 18 &  9825.81  &  5.9  &  2.1m   &    GG &   5720 - 7225 &\\
    &Apr 21 &  9828.78  &  5.9  &  2.1m   &    GG &   6280 - 8400 &\\
    &May 15 &  9852.85  &  5.9  &  2.1m   &    GG &   5760 - 7300 &\\
    &May 18 &  9855.82  &  5.9  &  2.7m   &    JT &              &\\
    &May 21 &  9858.80  &  5.9  &  2.1m   &    VW  &   3930 - 4275&\\
    &May 22 &  9859.70  &  5.9  &  2.1m   &    VW  & 3930 - 4275 &\\
    &Jun 10 &  9878.64  &  5.9  &  2.1m   &    GG  &   5720 - 7225 &\\
    &Jun 17 &  9885.64  &  5.9  &  2.7m   &    JT  &                &\\
    &Jun 19 &  9887.68  &  5.9  &  2.7m   &    JT  &                &\\
    &Jun 23 &  9891.69  &  5.9  &  2.1m   &    GG  &  5760 - 7230    &\\
    &Aug  7 &  9936.61  &  5.9  &  2.7m   &    JT  &                &\\
    &Aug  8 &  9937.62  &  5.9  &  2.7m   &    JT  &                &\\
    &Aug  9 &  9938.61  &  5.9  &  2.7m   &    JT  &                &\\
    &Sep 30  &  9990.62 &   6.1  &  2.7m  &   JT   &            &        \\
     &Oct  2  &  9992.64 &   6.3  &  2.7m  &   JT    &           &     \\   
      &Oct  7  &  9997.58 &   6.8  &  2.1m  &   REL   &    4880 - 5650&\\
      &Oct  8  &  9998.56 &   6.9  &  2.1m  &   REL   &    4880 - 5650 &\\
      &Oct  9  &  9999.56 &   7.0  &  2.1m  &   REL   &    5720 - 7225  &\\
      &Oct 11  &  10001.56 &   7.1  &  2.1m  &   REL   &    5720 - 7225  &\\
      &Oct 12  & 10002.55 &   7.3  &  2.1m  &   REL   &    6550 - 8550     &\\  
      &Oct 13  & 10003.59 &   7.7  &  2.7m  &   JT    &                    &\\
      &Oct 14  &  10004.56 &   8.0  &  2.7m  &   JT    &                 &\\
      &Oct 15  &  10005.55 &   8.4  &  2.7m  &   JT    &                  &\\
      &Oct 16  &  10006.61 &   8.8  &  2.1m  &   KR    &    4460 - 5040      &\\
      &Oct 17  &  10007.56 &   9.6  &  2.1m  &   KR    &    4460 - 5040    &\\
      &Oct 18  &  10008.56 &  10.1  &  2.1m  &   KR    &    4460 - 5040   &\\
      &Oct 18  &  10008.56 &  10.1  &  2.7m  &   DD    &        &         \\
      &Nov  2  &  10023.54 &  12.2  &  2.7m  &   CRJ   &        &         \\
      &Nov 12  &  10033.54 &  13.4  &  2.1m  &   CJK   &    5760 - 7225&\\     
      &Nov 14  &  10035.54 &  13.5  &  2.1m  &   CJK   &    5760 - 7225  &\\
      &Nov 15  &  10036.54 &  13.5  &  2.1m  &   CJK   &    5760 - 7225 &\\ \hline
\end{tabular}
\end{minipage}
\end{table*}

\begin{table*}
\setcounter{table} {0}
\centering
\begin{minipage} {140mm}
\caption { Catalogue of spectra of R\,CrB (continued).}
 \begin{tabular}{rlcccccr} \hline
 &   Date  &    JD - 2440000.0& {Mag.}\footnote{Visual magnitude from AAVSO. V (given to second decimal from Fernie (1997) and Efimov (1997).} &  Telescope & {Observer}%
 \footnote{Observers:  GG = Guillermo Gonzalez, AH = Artie Hatzes,
	     SLH = Suzanne Hawley, CRJ = C. Ren\'{e}e James, CJK = Chris Johns-
	     Krull, DLL = David L. Lambert, 
	     JT = Jocelyn Tomkin}
  &  {Comments}%
\footnote{This gives wavelength interval in \AA\ for the 2.1m spectra.} & \\ \hline 

  1996&Jan  5  &  10088.00 &  13.57 &  2.7m  &   JT    &        &         \\
      &Jan 19  &  10101.98 &  13.2  &  2.1m  &   GG    &    5570 - 6780&\\
      &Feb  1  &  10114.96 &  13.7  &  2.1m  &   CJK   &    5760 - 7225 &\\
      &Feb  6  &  10119.99 &  13.50 &  2.7m  &   JT    &                  &\\    
      &Feb  8  &  10122.01 &  13.6  &  2.7m  &   JT    &               &\\
      &Feb  9  &  10123.01 &  13.6  &  2.7m  &   JT    &         &        \\
      &Mar  2  &  10144.95 &  13.45 &  2.7m  & SLH/DLL &         &        \\
      &Mar 10  &  10152.99 &  13.5  &  2.1m  &   AH    &    5020 - 5910&\\
      &Mar 13  &  10155.90 &  13.4  &  2.1m  &   AH    &    5020 - 5910 &\\
      &Apr  9  &  10182.75 &  12.47 &  2.1m  &   GG    &    5510 - 6790&  \\
      &May  3  &  10206.90 &  10.8  &  2.7m  &   DLL   &                &\\
      &May  4  &  10207.88 &  10.7  &  2.7m  &   DLL   &                &\\
      &May  5  &  10208.85 &  10.40 &  2.7m  &   DLL   &        &         \\   
      &May  6  &  10209.85 &  10.5  &  2.7m  &   DLL   &              &\\
      &May  9  &  10212.82 &  10.07 &  2.1m  &   GG    &    5720 - 7220 &\\
      &May 31  &  10234.78 &   8.93 &  2.1m  &   CRJ   &    5720 - 7300  &\\
      &Jun  4  &  10238.77 &   8.88 &  2.1m  &   AH    &    5840 - 7360 &\\
      &Jun  5  &  10239.63 &   8.85 &  2.1m  &   AH    &    5990 - 7820 &\\
      &Jun 25  &  10259.72 &   8.1  &  2.1m  &   GG    &    5480 - 6780  &\\
      &Jul  8  &  10272.73 &   8.0  &  2.7m  &   CRJ   &      &\\
      &Jul 23  &  10287.65 &   7.4  &  2.7m  &   DLL   &   &\\
      &Jul 24  &  10288.61 &   7.5  &  2.7m  &   DLL   &     &\\
      &Jul 26  &  10290.64 &   7.5  &  2.7m  &   DLL   &    &\\
      &Oct  2  &  10358.55 &   7.0  &  2.7m  &   DLL   &    5880 - 5902&\\ \hline
\end{tabular}
\end{minipage}
\end{table*}

\begin{figure}
\epsfxsize=8truecm
\epsffile{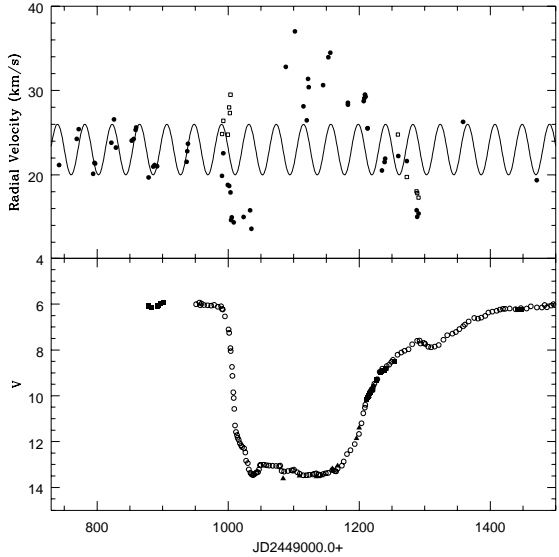}
\caption{Light curve and photospheric absorption line radial velocities
of R\,CrB during the 1995-1996 decline. The light curve (lower panel)
is constructed from visual observations 
kindly supplied by the AAVSO (open circles represent 10 day means)
 and V magnitudes from Efimov (1997 - filled triangles)
and Fernie (1997 - filled squares). Radial velocities (upper
panel) are presented for group A (dots) and B (open squares) lines (see text).} 
\end{figure}

At the 2.7m telescope, the
coud\'{e} cross-dispersed echelle 
spectrograph (Tull et al. 1995) was used with the
camera that gives a maximum (2-pixel)
resolving power of R =$\lambda/\Delta\lambda
= 60,000$. The detector
was a Tektronix 2048 $\times$ 2048 CCD. The recorded spectrum 
ran from about 3800\AA\ to 10000\AA, 
but spectral coverage was incomplete longward of
about 5500\AA. A Th-Ar hollow cathode lamp providing a wavelength calibration
 was observed either just prior
to or just after exposures of R CrB.
The pixel-to-pixel variation of the CCD was removed
using observations of a lamp providing a continuous spectrum. Typical exposure
times for R CrB in decline were 30 minutes with multiple exposures co-added
as necessary to improve the signal-to-noise ratio of the final spectrum.
Occasionally, an early-type rapidly-rotating star was observed to
provide a template of the telluric absorption lines. 

The Sandiford Cass\'{e}grain echelle spectrometer (McCarthy et al. 1993) was
used at the 2.1m telescope. These spectra have a resolving power also of
approximately R=60,000. Although the wavelength coverage at a single
exposure (see Table 1)
is less extensive than at the 2.7m,
orders of the echelle are completely recorded for wavelengths shorter than
about 7500\AA. Calibration procedures were the same as for the
2.7m telescope.

\begin{table}
\caption{Photometry from Efimov (1997) and Fernie (1997) of R CrB in the 1995 -- 1996 decline.}
\begin{center}

\begin{tabular}{rlccccc}
& & & & & & \\ \hline
  &   Date & U & B & V & R & I  \\ \hline

1995 & Oct 18 & 10.0 & 10.5 & 10.1 & 9.6 & 9.0 \\
     & Nov 2 &  12.59& 12.79& 12.02&11.52& ...  \\
1996 & Jan 5 & 13.67 & 14.05& 13.57&13.05&12.07 \\
     & Feb 6 & 13.50 & 14.00& 13.50&13.02&11.75 \\
     & Mar 2 & 13.45 & 13.92& 13.45&12.78&11.40  \\
     & Apr 9 & 13.15 & 13.35& 12.47&11.50&10.38 \\
     & May 5 & 12.13 & 11.67& 10.40& 9.50& 8.68 \\
\hline

\end{tabular}
\end{center}
\end{table}

Our spectra were not flux calibrated at the telescope. An adequate
calibration is possible using  sets of observed UBVRI magnitudes
(kindly supplied by Yu. S. Efimov and by J. Fernie)  in which we interpolate
to the dates of our observations. For spectra obtained early in the
decline UBVRI magnitudes are unavailable. In these cases, we identified
the visual magnitude  as the V magnitude. Colors
for these early observations were taken from UBVRI measurements on
earlier declines of R CrB with a similar rate of decline. The range
in colours from one decline to another is small and not a major source
of uncertainty.
 Adopted UBVRI
magnitudes are given in Table 2 for selected dates. Fluxes were computed from
these magnitudes using Wamsteker's (1981) calibration.
Clayton et al. (1997) published a flux calibrated low resolution spectrum taken
1996 April 7. Our derived fluxes for 1996 April 9 are in good agreement
with these published values.

\section{Maximum Light}

R CrB is known to be variable  at
maximum light.  
Photometric monitoring of R CrB  principally by Fernie and colleagues is
providing ample evidence of a continuous quasi-regular variation
in light (Fernie 1989, 1991, 1995, 1997; Fernie \& Seager 1994).
Variability  of the absorption line spectrum was
 detected long ago (Espin 1890).
 Recent observations at 
low (Clayton et al. 1995) and high spectral resolution (Rao \& Lambert 1997)
have begun to detail the changes.   

\subsection{Photospheric Radial Velocity}

 Our measurements of the photospheric radial velocity are based on a 
selection of 20 to 30  lines that we divide into two groups. Group A
comprises high-excitation {\it weak} lines of  N\,{\sc i},
 O\,{\sc i}, Al\,{\sc ii}, and S\,{\sc i}.
Group B is made up of {\it strong} lines of C\,{\sc i}, O\,{\sc i},
Si\,{\sc ii}, Ca\,{\sc i}, K\,{\sc i}, Cr\,{\sc ii},
and Ba\,{\sc ii}. 
Velocities are derived from the central core of a
line. Table 3  and Fig. 1 summarize these measurements
covering 558 days from about 8 months
before the decline  to near complete recovery to
maximum light.  

At maximum light,
the mean velocity of
22.5 $\pm 2.0$ km s$^{-1}$ and the range of 6 km s$^{-1}$
 from observations made between 1995 January and 1995
August
are the expected values for the star based upon
earlier studies (cf. Raveendran, Ashoka \& Rao 1986; Fernie \& Lawson 1993;  
Rao \& Lambert 1997). 
Throughout this interval, group A and B lines give the same velocity.

The historical data on R\,CrB's radial velocity were searched for a dominant
period. We  
collated radial velocity measurements 
based on spectra of coud\'{e} dispersion
- see Keenan \& Greenstein (1963),
Rao (1974), Fernie et al. (1972),
Gorynya et al. (1992), Fernie \& Lawson (1993), and Rao \& Lambert (1997).
The set comprises 149 measurements
from 1942 to 1995 when the star was not in decline.
The dominant source of velocity variations
is, of course, the atmospheric pulsation.
 Our periodogram
analysis  indicated a  pulsation period near
42.7 days. Experimentation with periods around this value suggest
that a period of 42.6968 days, a mean velocity of 22.5 km s$^{-1}$,
and a range of 6 km s$^{-1}$  provides the best fit to the measurements
over the half-century.
 Several investigators 
have mentioned that the pulsational period
is not strictly constant;  a particular value may represent the 
photometric data  for an interval of one to two years. 
Slight variations of this period or small phase shifts
 seem to be indicated. For example, if
 the earliest measurements assembled by
Keenan \& Greenstein (1963) are dropped, the best-fitting period
is lengthened slightly to 42.7588 days.

 In
Fig. 1, we show the measured velocities: filled circles denote either
 the mean velocity of group A and B lines where there is no significant
difference between the two groups or the velocity of the A lines where there
is a significant difference, and the open squares denote group B velocities
where they differ from the A velocities by more than 2 km s$^{-1}$.
The line is the `historical'
sine curve with a period of 42.6968 days and 
a range of 6 km s$^{-1}$ around a mean velocity of 22.5 km s$^{-1}$. (The
difference between a period of 42.6968 and of 42.7588 days is unimportant
over this short interval of time.) 
Observations prior to the onset of the decline  are
closely matched by the sine curve.

 Both group A and B lines
depart in different ways from this curve at the onset
but the difference between group A and B lines disappears after a few days.
Then,
 the radial
 velocity shown by photospheric (group A and B) lines is systematically
more positive throughout the deepest part of the decline
 than predicted by the sine curve.
These deviations occur at
a time when the absorption lines have very unusual profiles. Lines unaffected
by emission show shallow asymmetric profiles quite unlike photospheric
profiles seen at maximum light. 
These changes and the marked redshift
are attributed to scattering of photospheric light by R\,CrB's dusty
envelope (Sec. 9.3). We presume that the obscured photosphere pulsated
throughout according to the sine curve shown in Fig. 1.  
 The two measurements from late in the recovery 
match well the predicted sine curve showing that the pulsation after the
decline followed the ephemeris that matched the pre-decline observations. 
These observations indicate that, except for the photospheric disturbance
at the onset (Sec. 4), the photosphere pulsated oblivious to the
cloud of soot obscuring it from our view.


\begin{table*} 
\centering
\begin{minipage}{140mm}
\caption{Radial velocities of photospheric absorption lines.} 
\begin{tabular}{rccccccc} \hline 
\multicolumn{1}{c}{Date} & \multicolumn{1}{c}{JD-2440000}& 
\multicolumn{6}{c}{Line Selection%
\footnote{Velocity V is given in km s$^{-1}$ followed by the number of
lines n. }}
\\ \cline{3-8}
 && \multicolumn{2}{c}{All}&\multicolumn{2}{c}{Group A}& 
\multicolumn{2}{c}{Group B}\\ \cline{3-8}
 && \multicolumn{1}{c}{V}&\multicolumn{1}{c}{n}&\multicolumn{1}{c}{V}& 
\multicolumn{1}{c}{n}&\multicolumn{1}{c}{V}&\multicolumn{1}{c}{n}\\ \hline
    1995 Jan 24 & 9742.01   &   21.2  &   22  & 21.0    &  9  & 21.3 & 13\\
     Feb 20 & 9768.97   &   24.3  &   16  & 24.0   &   9  & 24.7 &  7\\ 
     Feb 23 & 9771.92   &   25.4  &   17  & 25.1   &   9  & 25.8 &  8\\
     Mar 17 & 9793.96   &   20.1  &   17  & 19.8   &   9  & 20.5 &  8\\
     Mar 19 & 9796.02   &   21.4  &   15  & 20.4   &   7  & 22.3 &  8\\
     Mar 20 & 9797.00   &   21.3  &   14  & 20.3   &   7  & 22.4 &  7\\
     Apr 14 & 9821.80   &   23.8  &   22  & 22.5   &   9  & 24.7 & 13\\
     Apr 18 & 9825.81   &   26.6  &   24  & 25.1   &  11  & 28.0 & 13\\
     Apr 21 & 9828.78   &   23.2  &   22  & 21.6   &  10  & 24.6 & 12\\
     May 15 & 9852.85   &   24.1  &   24  & 22.5   &  10  & 25.0 & 14\\
     May 18 & 9855.82   &   24.3  &   25  & 23.7   &  10  & 24.7 & 10\\
     May 21 & 9858.80   &   25.3  &    3  & ...    & ...  & 25.3 &  3\\
     May 22 & 9859.70   &   25.6  &    2  & ...    & ...  & 25.6 &  2\\
     Jun 10 & 9878.64   &   19.7  &   21  & 19.2   &  21  & 20.0 & 13\\
     Jun 17 & 9885.64   &   21.0  &   27  & 20.2   &  10  & 21.4 & 17\\
     Jun 19 & 9887.68   &   21.2  &   26  & 20.6   &  12  & 21.4 & 14\\
     Jun 23 & 9891.69   &   21.0  &   21  & 19.8   &  10  & 21.8 & 11\\
     Aug  7 & 9936.61   &   21.5  &   26  &  20.9  &  12  & 22.0 & 14\\
     Aug  8 & 9937.62   &   22.8  &   26  &  22.0  &  11  & 23.0 & 15\\
     Aug  9 & 9938.61   &   23.7  &   29  &  23.2  &  13  & ...   &...\\
     Sep 30 & 9990.62   & ...     & ...   &  19.9  &  13  & 24.8  & 16\\
     Oct  2 & 9992.64   & ...     & ...   &  22.6  &  12  & 26.4  & 16\\
     Oct  7 & 9997.57   & ...     & ...   &  ...   &  ... & ...   &...\\
     Oct  8 & 9998.55   & ...     & ...   &  ...   &  ... & ...   &...\\
     Oct  9 & 9999.56   & ...     & ...   &  18.8  &   7  & 24.8  & 10 \\   
     Oct 11 & 10001.55   &...      &...    &  18.7  &   9  & 27.3  & 11  \\
     Oct 12 & 10002.55   &...      &...    &  19.4  &   9  & 29.9  &  7   \\
     Oct 13 & 10003.59   &   22.9  &   26  &  17.9  &  13  & 29.5  & 12   \\
     Oct 14 & 10004.56   &   14.9  &   16  &  14.6  &  14  & 17.0  &  2 \\
     Oct 15 & 10005.55   &   15.0  &   16  &  14.8  &  14  & 15.9  & 2\\
     Oct 18 & 10008.56   &   14.4  &   14  &  14.0  &  13  & 13.5  & 1  \\
     Nov  2 & 10023.54   &   15.0  &   15  &  16.0  &  11  & 14.0  & 4 \\
     Nov 12 & 10033.54   &   15.8  &    5  &  15.3  &   2  & 16.9  &3\\
     Nov 14 & 10035.54   &   13.6  &    2  &  ...   &  ... & 13.6  &2 \\
\hline
\end{tabular}
\end{minipage}
\end{table*}

\begin{table*} 
\setcounter {table}{2}
\centering
\begin{minipage}{140mm}
\caption{Radial velocities of photospheric absorption lines (continued).} 
\begin{tabular}{rccccccc} \hline 
\multicolumn{1}{c}{Date} & \multicolumn{1}{c}{JD-2440000}& 
\multicolumn{6}{c}{Line Selection%
\footnote{Velocity V is given in km s$^{-1}$ followed by the number of
lines n. }}
\\ \cline{3-8}
 && \multicolumn{2}{c}{All}&\multicolumn{2}{c}{Group A}& 
\multicolumn{2}{c}{Group B}\\ \cline{3-8}
 && \multicolumn{1}{c}{V}&\multicolumn{1}{c}{n}&\multicolumn{1}{c}{V}& 
\multicolumn{1}{c}{n}&\multicolumn{1}{c}{V}&\multicolumn{1}{c}{n}\\ \hline
    1996 Jan  5 & 10088.00   &   32.8  &   13  &  33.6  &   5  & 32.3  & 8\\
     Jan 19 & 10101.97   &   37.0  &    5  &  35.7  &   2  & 37.9  &3\\
     Feb  1 & 10114.96   &   28.1  &    5  &  26.9  &   4  & 27.8  &1\\
     Feb  6 & 10119.99   &   26.5  &    9  &  26.5  &   9  & ...   &...\\
     Feb  8 & 10122.01   &   31.4  &    8  &  31.8  &   3  & 31.2  &5\\
     Feb  9 & 10123.01   &   30.4  &    8  &  29.7  &   4  & 30.4  &4\\
     Mar  2 & 10144.95   &   30.6  &   10  &  30.8  &   7  & 29.9  &3\\
     Mar 10 & 10152.99   &   34.0  &    3  &  ...   &  ... & 34.0  &3\\
     Mar 13 & 10155.90   &   34.5  &    2  &  ...   &  ... & 34.5  &2\\
     Apr  9 & 10182.74   &   28.5  &   10  &  28.6  &   5  & 27.9  &5\\
     Apr  9 & 10182.77   &   28.3  &   13  &  27.2  &   5  & 29.3  &8\\
     May  3 & 10206.90   &   28.7  &   13  &  28.5  &   8  & 29.0  &5\\
     May  4 & 10207.88   &   29.1  &   19  &  28.4  &  10  & 29.9  & 10    \\
     May  5 & 10208.85   &   29.5  &   19  &  29.2  &   9  & 29.8  & 10   \\
     May  6 & 10209.85   &   29.3  &   19  &  29.0  &   9  & 29.5  & 10  \\
     May  9 & 10212.82   &   25.5  &   13  &  24.5  &   7  & 26.7  & 6   \\
     May  9 & 10212.83   &   25.5  &   19  &  24.4  &   7  & 26.2  & 12   \\
     May 31 & 10234.78   &   20.5  &   19  &  21.1  &   7  & 20.2  &12\\
     Jun  4 & 10238.77   &   21.5  &   19  &  21.6  &   7  & 21.4  &12\\
     Jun  5 & 10239.63   &   21.9  &   24  &  21.9  &  10  & 21.9  &14\\
     Jun 25 & 10259.72   &   23.1  &   13  &  21.9  &   7  & 24.8  &6\\
     Jul  8 & 10272.73   &   20.3  &   22  &  21.6  &   7  & 19.7  &15\\
     Jul 23 & 10287.65   &   17.1  &   25  &  15.8  &  10  & 18.1  &15\\
     Jul 24 & 10288.61   &   16.4  &   22  &  15.0  &  11  & 17.8  &11\\
     Jul 26 & 10290.64   &   16.0  &   14  &  15.4  &   8  & 17.3  &6\\  \hline 
\end{tabular}
\end{minipage}
\end{table*}

\subsection{Sharp Emission Lines at Maximum light}

At maximum light, strong  low excitation lines of neutral atoms and 
singly-charged ions show an
apparent doubling in their absorption cores, as first seen by
Payne-Gaposchkin (1963),
and Keenan \& Greenstein (1963) and confirmed 
from  high-resolution CCD spectra by Lambert et al. (1990).
 The doubling is considered to
result from the superposition of an emission component on the
photospheric absorption core.
Our present spectra confirm that the
emission is probably a permanent feature at maximum light.
An excellent spectrum obtained on 1995 May 18
clearly shows emission in Sc\,{\sc ii} 4246\AA, Sr\,{\sc ii} 4077 and 4215\AA,
the Na\,{\sc i}  D lines, and the Ca\,{\sc ii} infrared triplet lines, representing
the emission spectrum E2. The
emission is at a velocity of 18 $\pm$ 1 km s$^{-1}$  or shifted to the
blue by about 5 km s$^{-1}$ relative to the photosphere's systemic velocity.
Fig. 2 shows the  Sc\,{\sc ii} 4246\AA\ line on three occasions in 1995 prior to
the decline, two occasions right at the onset of the decline, and when the
star had faded by about 1.6 magnitudes.
 The emission
core is present prior to onset with an intensity that appears slightly variable
but this variation may reflect a varying continuum flux resulting from
the pulsation.
Emission at the same velocity is striking in the 1995 October 13 spectrum.
Continuing the sequence,
Fig. 3 shows the May 18 maximum light spectrum, the October 13 spectrum,
and the October 18 spectrum when the star had faded by 4 magnitudes. On this
latter spectrum, weaker emission is clearly present in all the
photospheric lines in this region. Incipient emission is present
affecting these lines in the October 13 spectrum shown in Figure 2. 

Our inference is that sharp emission lines are a permanent presence.
As we show below, these sharp emission lines are of constant velocity,
 unreddened, and of 
constant flux  until
the photosphere is dimmed by about 4 magnitudes. These are surely clues
to the location of the lines' emitting region.

\begin{figure}
\epsfxsize=8truecm
\epsffile{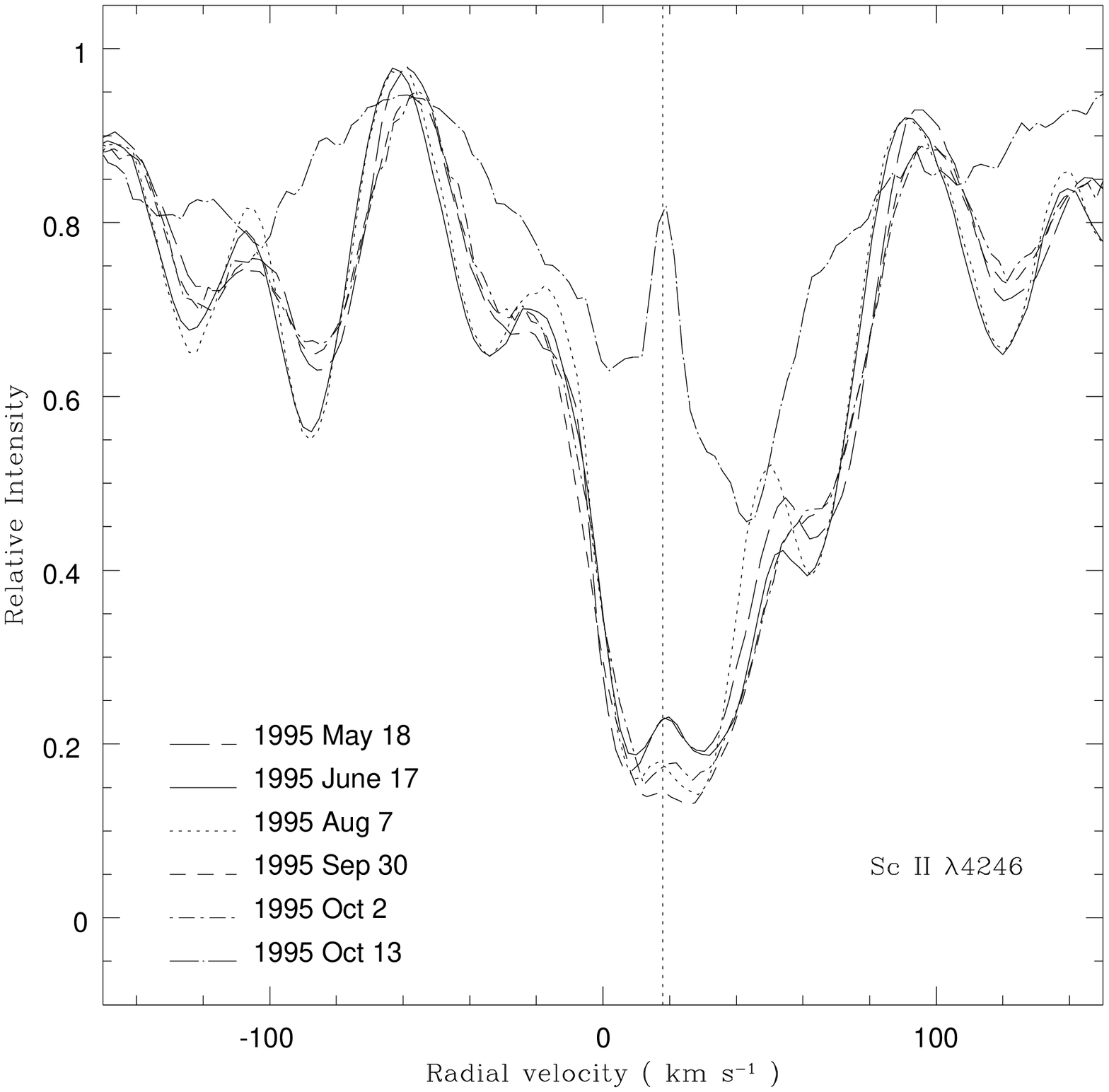}
\caption{The
central emission core of the Sc\,{\sc ii}  4246\AA\ line on
5 occasions prior to the 1995 decline, and on  1995 October 13  when the
star had faded by 1.6 magnitudes.  The vertical broken line denotes 
a radial velocity of 18 km s$^{-1}$.} 
\end{figure}

\begin{figure}
\epsfxsize=8truecm
\epsffile{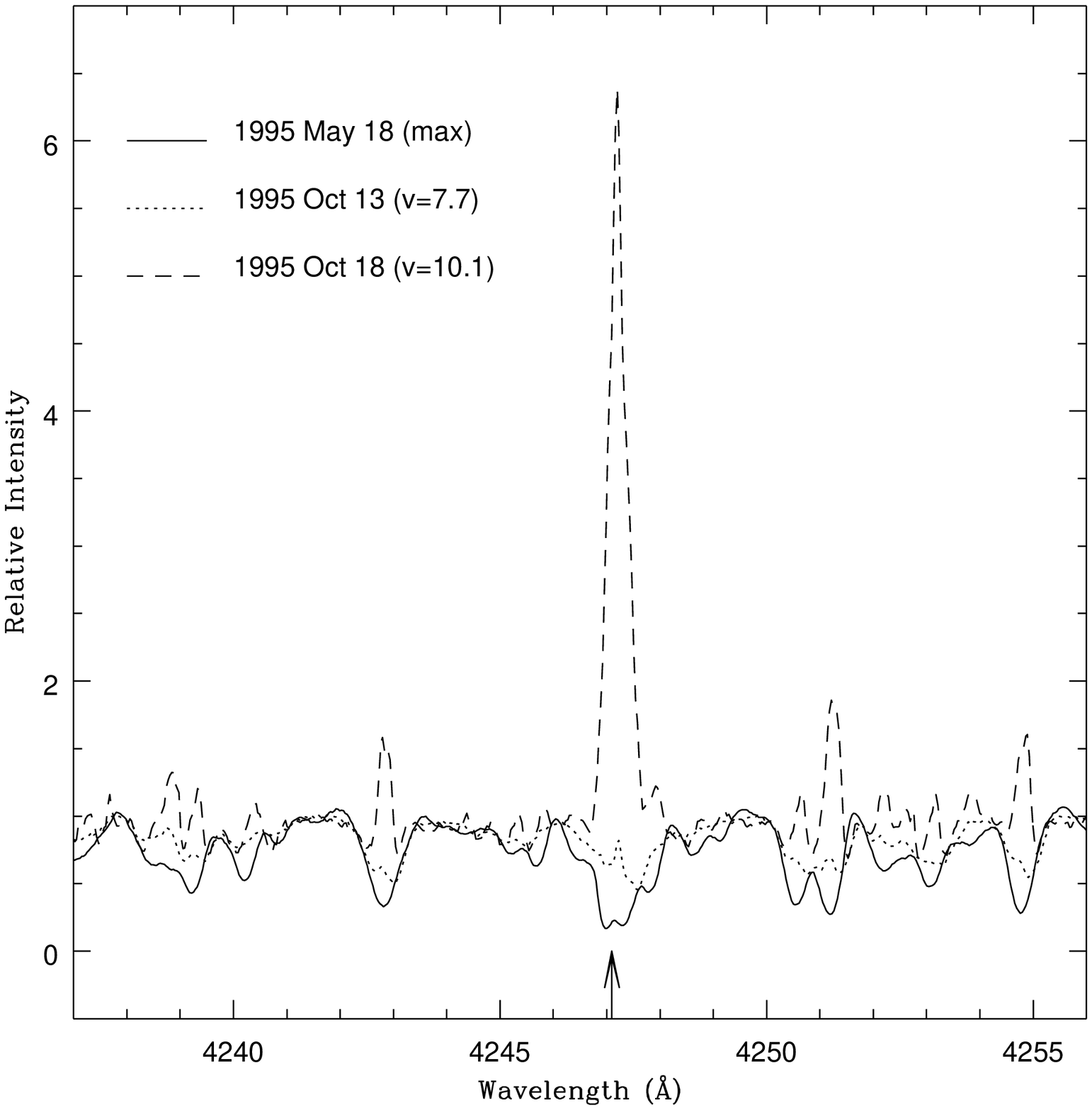}
\caption{The emergence of the sharp emission line spectrum near the
Sc\,{\sc ii} 
4246\AA. The key indicates
when these spectra were obtained and the visual magnitude
on those dates.} 
\end{figure}

\section{The Onset of the Decline - Photospheric Activity}

Our spectra reveal remarkable changes in high-excitation photospheric lines
over a fortnight's interval beginning with the onset of the
decline. 
 It seems
probable that the changes  betray information about the 
mysterious trigger of an RCB decline. In particular, this discovery
sites the trigger in the star's photosphere and eliminates some
ideas about the cause of  the decline: i.e., the decline is not
brought on by passage of a circumstellar
cloud across the disk of the star, or by spontaneous condensation
of soot in the cool outer reaches of the atmosphere.

The changes are well illustrated in Figure 4 showing  comparisons of
the spectra obtained
on 1995 September 30 and 1995 August 9.
On September 30 R CrB remained close to maximum light but by October 2,
the date on which the next spectrum was acquired, it had begun to
fade. Beginning with the September 30 spectrum, there is a clear
velocity difference (Table 3, Fig. 1) between the group A and B
photospheric lines, as measured from their line cores.
 Unfortunately, it is not possible to date precisely
 the onset
of this velocity difference except to note that it was not present
from August 7 to 9. The difference increased almost monotonically from about
5 km s$^{-1}$ on 1995 September 30 to about 12 km s$^{-1}$ by 1995
October 13 when the star had faded to V $\simeq$ 7.7.
Since the  group B lines are  formed closer to the surface than
the group A lines, we infer that  shallower photospheric layers were 
falling in towards the deepest visible layers at velocities in excess of the
sound speed ( $\simeq 5$ km s$^{-1}$).
 Shortly after October 13,
emission appears in the  core of a group B line (Fig. 4). 
Transition from a broader than usual absorption line to an emission line in the
wing of an absorption line occurred between 1995 October 9 and 13. Emission
persisted to just prior to 1995 November 2 (V$\simeq 12$) when the  profiles
of group B lines again resembled photospheric lines observed near
maximum light, and the
velocity difference between group A and B lines was less than 2 km s$^{-1}$
with a mean velocity  blue-shifted by about 8 km s$^{-1}$
relative to the systemic velocity. These emission lines belong to the
class of E1 lines.

\begin{figure}
\epsfxsize=8truecm
\epsffile{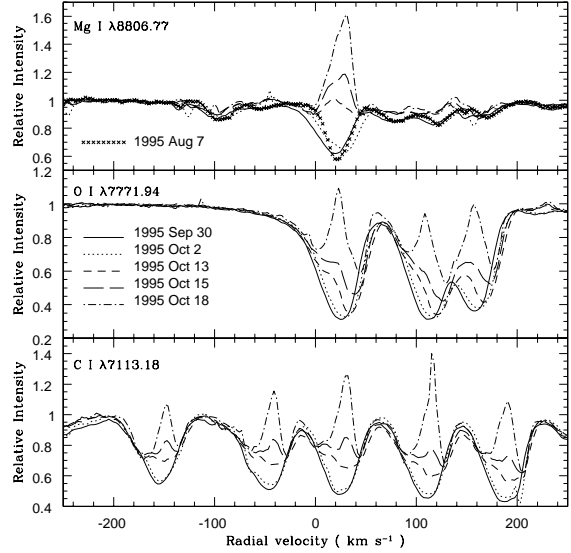}
\caption{Profiles of high-excitation lines from onset of the decline near
1995 September 30 through to 1995 October 18 when the R\,CrB had
faded to V$\simeq 10$. The top panel shows the Mg\,{\sc i} 8806\AA\ line.
The middle panel shows the O\,{\sc i} triplet with the velocity scale
set to the 7771.94\AA\ line. The bottom panel shows  C\,{\sc i} lines
near 7100\AA\ with the velocity scale set to the 7113.18\AA\ line.
 }
\end{figure}

 In late September, the
photospheric velocities according to the 42.6968 day sine curve
should have been close to their maximum value of about 26 km s$^{-1}$.
Lines of group B,
the strong to very strong lines of
C\,{\sc i}, O\,{\sc i},  and other species, are close
to the expected velocity but lines
of group A, high excitation weak lines of N\,{\sc i} and other species, are
blue-shifted with respect to the predicted maximum velocity.
Group B lines  
 on September 30 are 
are also broadened relative to their August 7 profiles and to
 the group A profiles
in this early phase of the decline. For all lines, the red wing 
remains at about the same velocity but there is 
 additional
absorption on the blue side of group B lines.
 In contrast, the group A
 profiles are largely unchanged, although shifted in velocity.
Although the group B lines later show an emission component, it is 
 unlikely that
these differences in profile 
 are initially or solely
 due to emission altering an underlying unchanging absorption
profile because the equivalent widths of the affected lines are
larger in September 30 than in August 9.
The profile changes during onset of the decline are considerably
more extreme than those occuring during  regular pulsations
at maximum light (Rao \& Lambert 1997).

These emission lines associated with the onset -- the `transient' or E1 lines --
 are to be distinguished from
the sharp emission (E2) lines  present in and out of a decline. 
A distinguishing feature of the onset-related or transient emission lines is
their large range of excitation potential. At the top end are lines
with  
lower excitation potential in the range 6.5 to 9.5 eV. Such lines
are {\it not} contributors of sharp (E2) lines. Lines of lower excitation
potential  are blends of a transient and a sharp line and  include  Li\,{\sc i},
Ca\,{\sc i}, Fe\,{\sc ii}, Ni\,{\sc i}, and La\,{\sc ii} transitions.  
Another notable contributor of transient but not sharp
lines is the C$_2$ Swan system
(see below). Significantly, group A lines do not
appear in emission. The fluxes of the C\,{\sc i} transient emission
lines  provide an estimate of the excitation temperature. The
strongest lines are optically thick in that lines from the same multiplet
do not scale with the gf-values of the lines. Lines that do appear
to be optically thin suggest an excitation temperature of about 8400K.
This temperature is definitely higher
than that estimated from the sharp emission lines.
 The velocity of the transient C\,{\sc i}
emission lines is 20 to 30 km s$^{-1}$, i.e, a velocity between that of the
group A absorption lines and the red-shifted absorption component 
accompanying transient emission lines.


On 1995 October 18 almost all lines showed a red absorption
component or an inverse-P Cygni profile.\footnote{Inverse P Cygni profiles for
low excitation lines were observed by Vanture \& Wallerstein (1995) on a
spectrum of RY\,Sgr taken during the recovery from its 1993 deep minimum.
High excitation lines were not seen in this spectrum.} Fig. 5 is a montage
of emission lines with the red-shifted absorption indicated. 
Many lines are a blend of transient and sharp emission components
at a very similar velocity.
  The red absorption component  is not 
a portion of the photospheric line not filled in by emission.
This assertion is based on the fact that the full array of
lines gives the same velocity for this component: 43 km s$^{-1}$ from 90
lines or a red shift of 30 km s$^{-1}$ with respect to the 
mean velocity of group A and B lines or 20 km s$^{-1}$ relative to the
photosphere's systemic velocity. 
If the red absorption were a residual of the photospheric
line, we would expect the apparent velocity to vary with the strength
of the overlying emission. More significantly, the depth of the red absorption
for many lines is deeper than the photospheric line depth at the same
velocity from the line core, as clearly seen
in the Ba\,{\sc ii} 5854\AA\ line (Fig. 6).
 The red-shifted absorption is also a transient
phenomenon and by 1995 November 2 had disappeared.

 Curiously,
 a few lines,
 e.g., Ni\,{\sc i} 7789 \AA\  (Fig. 7), exhibit a P Cygni profile
with the absorption component  at the photospheric
velocity.  Oddly,
P Cygni profiles appear restricted to a few Ni\,{\sc i} multiplets and
high multiplets of Fe\,{\sc i} (e.g., RMT1107).
The P Cygni profile was not seen on or after 1995 November 2.
The difference  between P Cygni and inverse P Cygni profiles
could be due to a velocity differential between
the layer providing the absorption line and that providing the emission
line.  Since the latter appears at about the same velocity for all lines,
the absorption line is shifting such that weak lines (e.g., Ni\,{\sc i}) formed
at the top of the photosphere are blue-shifted relative to lines formed
deeper in the atmosphere (e.g., O\,{\sc i}): the  
 velocity shifting to more positive velocities
for lines formed at shallower depths in the photosphere.

\begin{figure}
\epsfxsize=8truecm
\epsffile{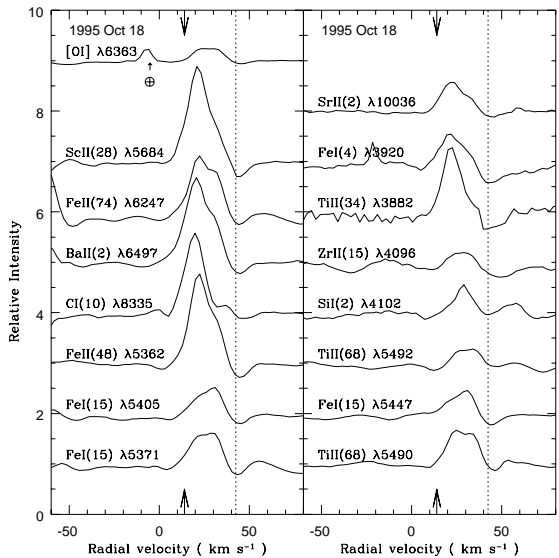}
\caption{
Selected  emission lines from the 1995 October 18 spectrum.
The photospheric velocity, as measured from lines without an obvious
emission component, is indicated by the arrow top and bottom of the figure.
Most lines have a red-shifted absorption component at 43 km s$^{-1}$ which
is indicated by the dotted line.}
\end{figure}

\begin{figure}
\epsfxsize=8truecm
\epsffile{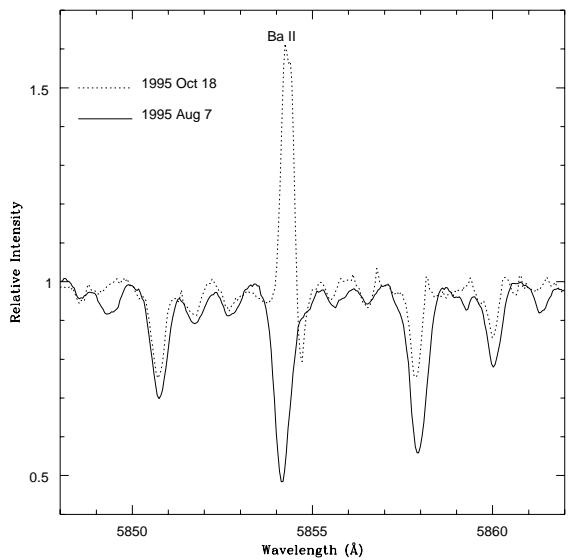}
\caption{
 The Ba\,{\sc ii} 5854\AA\ line on 1995 August 7 prior to the
decline and on 1995 October 18 when the star had faded by 4 magnitudes. Spectra
have been aligned so that the weaker photospheric lines are superimposed.
Note the depth of the  sharp absorption associated in the Ba\,{\sc ii} line's
red wing.}
\end{figure}

\begin{figure}
\epsfxsize=8truecm
\epsffile{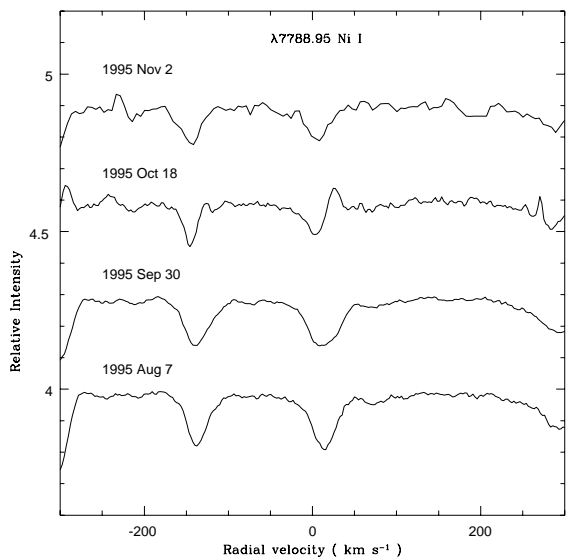}
\caption{ Evolution of the Ni\,{\sc i} line at 7789\AA\ from prior to the
decline through to 1995 November 2  when the star had faded by about 6
magnitudes. Note the P Cygni profile on 1995 October 18 and the disappearance
of emission by 1995 November 2.}
\end{figure}

An additional absorption component is seen in the strong Na\,{\sc i} D lines
following the onset of the decline.
In addition to  changes in their photospheric profiles, the
Na\,{\sc i} D lines  showed on 1995 September 30
additional absorption in their
blue wing (Fig. 8). (Similar changes  occurred in the Ca\,{\sc ii} infrared
triplet lines.) This absorption which appears to be a narrow component is at
about -3 km s$^{-1}$ or blue-shifted by nearly 30 km s$^{-1}$
relative to the anticipated photospheric velocity.

\begin{figure}
\epsfxsize=8truecm
\epsffile{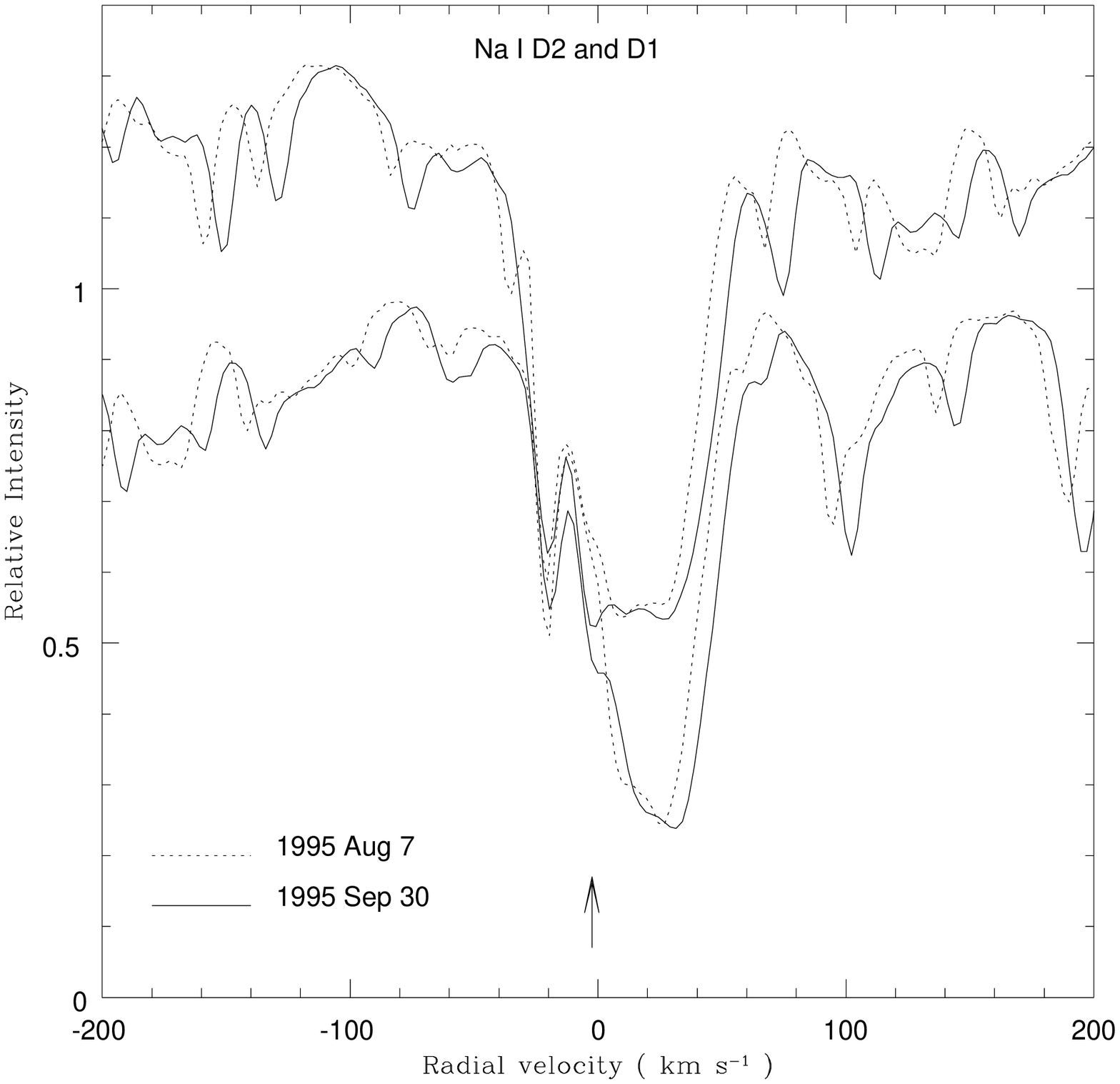}
\caption{Profiles
of the Na\,{\sc i} D$_1$ and D$_2$ lines  on 1995 August 7 and 1995
September 30. The August 7 spectrum was obtained at maximum light.
 Additional absorption 
(marked by the arrow) appears in the September 30 spectrum when the
decline was just beginning. The sharper
component to the blue is a permanent unchanging interstellar (or circumstellar)
pair of unresolved lines. Spectra have been aligned such that this
interstellar component is at its heliocentric velocity.
 Weak sharp lines that seem to be different in the
two spectra are telluric H$_2$O lines.}
\end{figure}

Our spectra show that 
membership in the E1 class of emission lines
must be extended to include high-excitation lines of C\,{\sc i},
O\,{\sc i}, and other species.\footnote{Some earlier reports based on
 photographic spectra
did note a filling in of C\,{\sc i} lines in the blue but red lines were
not observed photographically. Our spectra  show transient weak
emission cores in the stronger C\,{\sc i} lines in the blue (e.g., RMT6 at
4762-4776\AA).} The primary factor behind the evolution of the E1 spectrum
is the changing physical conditions in the emitting regions and not
the occultation of these regions by the developing dust cloud. 
The mix of line profiles from P Cygni
to inverse P Cygni is difficult to understand if occultation by
remote dust is dominant but more readily understood if the transient lines
 are emitted by atmospheric layers experiencing shocks. 
Emission in the high-excitation lines lasts a brief while. At its
disappearance, the absorption profiles are returned to their pre-decline
condition (see below) because, we suggest, the disturbed atmospheric
layers have relaxed to approximately their normal state.   If occultation
by the fresh dust cloud were to control the transient lines' appearance,
it would be necessary to suppose that the optical depth to the photosphere
were less than that to the lines' emitting region. This seems unlikely.
Moreover, shock excited C\,{\sc i} emissions are present in RY Sgr during its
pulsation cycle (Cottrell \& Lambert, unpublished observations)
 at maximum light, similar to the
emissions seen above.
If future observations show that behaviour of high-excitation lines in 1995
was typical of all declines, an explanation in terms of occultation by
dust will be excludable. 

\section{R CrB around Minimum Light -- The Sharp Emission Line Spectrum}

\subsection{Introduction}

After 1995 November 2 (V = 12.2), the emission line spectrum comprises low
excitation lines of mainly singly-ionized metals.
 Representative line
profiles are shown in Fig. 5  for 1995 October 18: the C\,{\sc i}
 8335\AA\ line
is a transient line, but lines such as  Fe\,{\sc i} 5405 and 5371\AA\
  are 
sharp lines.  Evolution of the emission line spectrum is shown by
 Fig. 9 and 10. 

\begin{figure}
\epsfxsize=8truecm
\epsffile{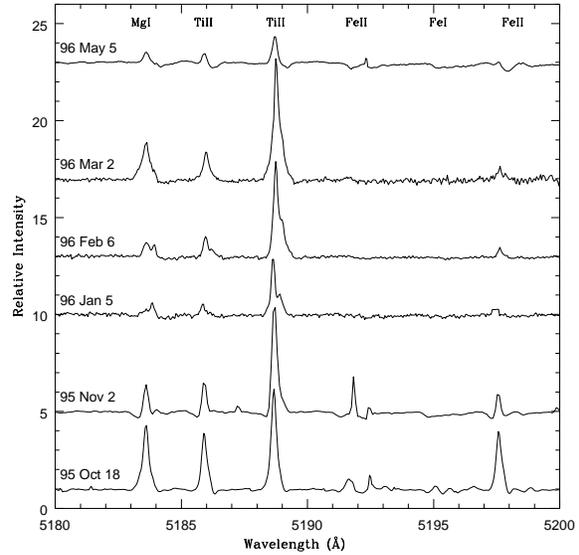}
\caption{Sharp emission lines on representative spectra from
early in the decline to late in the recovery.}
\end{figure} 

\begin{figure}
\epsfxsize=8truecm
\epsffile{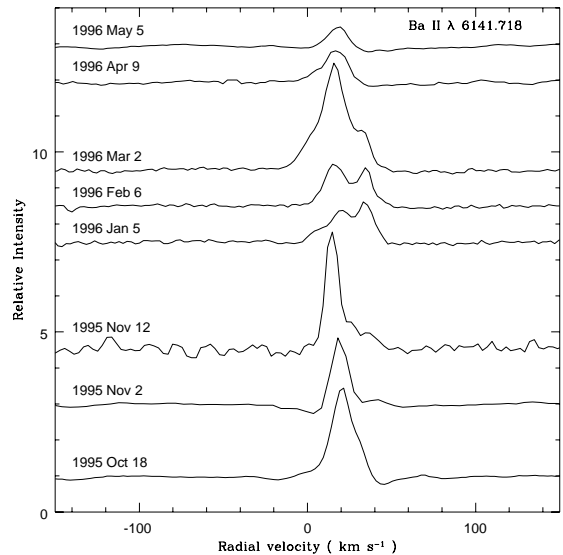}
\caption{
Evolution of the Ba\,{\sc ii} 6142\AA\ line from early in the decline to
late in the recovery. }
\end{figure}

The sharp emission lines appear
 composed of
two or three components which we label C1, C2, and C3 in order of increasing
velocity. 
The lines are  resolved; instrumental (FWHM) width as measured
from the Th comparison lines is about 5 km s$^{-1}$ but the C2
component 
has a FWHM of about  14 km s$^{-1}$. The emission lines
are not broader than the same lines in absorption at
maximum light: the base width
of the emission and absorption lines are both about 40 km s$^{-1}$.
The principal component (C2) 
is at 20 km s$^{-1}$ or displaced by - 3  km s$^{-1}$ from the
systemic velocity.
This shift is slightly smaller than  reported at earlier declines
for R\,CrB and RY\,Sgr.

\subsection{Forbidden Lines}

In all previous discussions of the sharp emission line spectrum
of R\,CrB stars in decline, identified lines were exclusively
permitted lines. Indeed, permitted lines comprise the vast
majority of sharp  lines in our spectra.
 Searches for forbidden lines, where reported, were 
described as unsuccessful.
Since forbidden lines  
may provide data on 
physical conditions in the emitting gas,  we searched for a variety
of forbidden lines.

We  have identified sharp
forbidden lines for the first time. Identifications include
[C\,{\sc i}] 8727, 9823, and 9850\AA,
[O\,{\sc i}] 5577, 6300, and 6363 \AA, and
[Ca\,{\sc ii}] 7291 and 7323\AA.\footnote{[Ca\,{\sc ii}]
lines
had been identified previously in the 1977 decline of R\,CrB
 (Herbig 1990)
and in RY Sgr by Asplund (1995) but the spectral resolution did not permit
a clear differentiation between sharp and broad emission in  these
lines. The [O\,{\sc ii}] 3727\AA\ must have been present but our
spectra have too low a S/N ratio at that wavelength. },
 and  several  detections of
[Fe\,{\sc ii}] lines. Forbidden lines
have the profile (component structure) and the velocity of the permitted
sharp lines.

{\bf [C\,{\sc i}]}. The 9850\AA\ [C\,{\sc i}] line was
 seen first  on 1995 October 13 and was last seen
on 1996 May 5. 
The excited 8727\AA\ line was present on 1995 October 13 but unfortunately
later spectra did not include this wavelength region. 
The estimated flux ratio for 1995 October 13 is F(8727)/ [F(9850) +
F(9823)] = 2.4 $\pm$  0.3 where the contribution of the 9823\AA\ line is
estimated from that of 9850\AA\ and the known branching ratio.

{\bf [O\,{\sc i}]}. The  lines 6300, 6363, and 5577\AA\ are
present. The 6363\AA\ line (Fig. 5)
 was first seen on 1995 October 15, and like
the [C~I] 9850\AA\ line was present throughout the decline.
We estimate that the flux  ratio 
[F(6300) + F(6363)]/f[5577] $\sim 18$ throughout the decline.

{\bf [Ca}\,{\sc ii}]. The 7291\AA\ and 7323\AA\ lines are the strongest
forbidden lines that are sharp (Fig. 11).
 Strong sharp components
 seen for the two forbidden lines and the
 permitted infra-red triplet are superimposed on weak broad
emission lines. 
Relative
fluxes in the  forbidden and the permitted infra-red and H/K lines 
 are discussed below.

\begin{figure}
\epsfxsize=8truecm
\epsffile{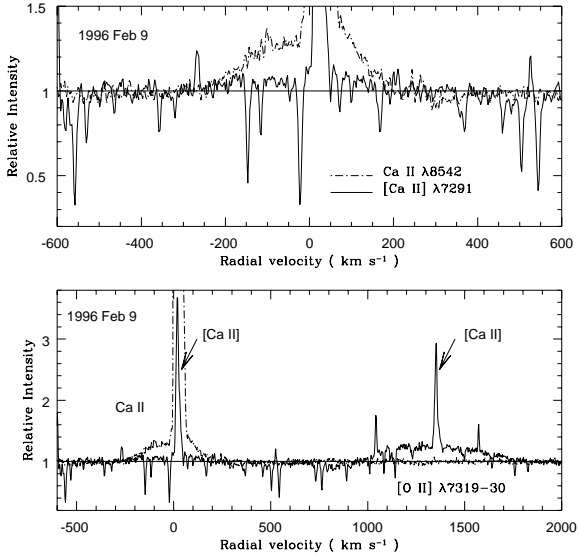}
\caption{The [Ca\,{\sc ii}] lines at 7291 and 7323\AA. The top panel shows the 
7291\AA\ line and the infra-red triplet 8542\AA\ line. The latter line has a
strong sharp component not fully shown and a broad component. The
[Ca\,{\sc ii}] 7291\AA\ line has a sharp component
 also not fully shown and a hint
of a broad component. Sharp lines across the 7291\AA\ spectrum are telluric
H$_2$O lines. The lower panel for which the velocity scale is set to the
7291\AA\ line's rest wavelength of 7291.46\AA\ shows both forbidden lines.
The broad emission around the [Ca II] 7323\AA\ line is a blend of [O II]
lines.}
\end{figure} 

{\bf [Fe\,{\sc ii}]}. Unsuccessful searches for [Fe\,{\sc ii}]
 lines were reported from photographic spectra
by Herbig (1949) and Payne-Gaposchkin (1963) 
for R\,CrB in decline, and by Alexander et al. (1972) 
for RY\,Sgr. Our
spectra show weak [Fe\,{\sc ii}] emission lines from 1995 October 18 to 1996 
February 6.
  Fig. 12  shows three lines from the
1996 February 6 spectrum. Radial velocities of the lines coincide
with that of the central component of the permitted lines.  The
flux ratio of forbidden to permitted lines evolved with the
former becoming relatively stronger until about the middle of the
deepest part of the decline. For example, the flux ratio of the
4244\AA\ [Fe\,II] to the 4233\AA\ Fe\,II lines increased from 0.008
on 1995 October 18, to 0.13 on 1995 November 13, and to 0.45 on 1996
February 6.

\begin{figure}
\epsfxsize=8truecm
\epsffile{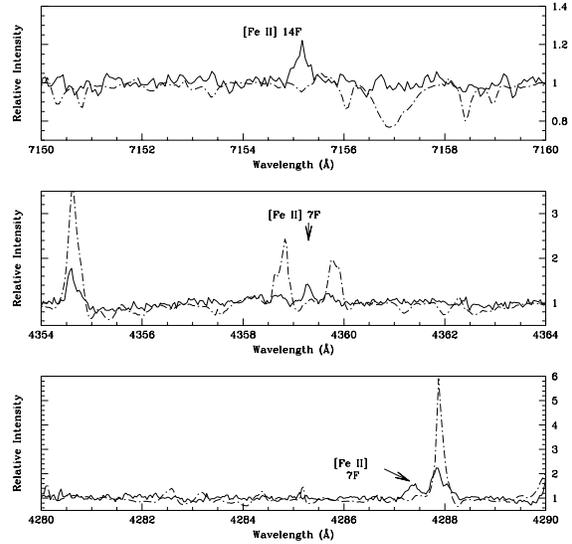}
\caption{[Fe\,{\sc ii}] lines in the spectrum of R CrB at minimum light.
 [Fe\,{\sc ii}]
lines present in the 1996 February 6 spectrum (solid line) are identified.
The comparison spectrum for 1995 October 18 (dash-dot line) does not
show these forbidden lines.}
\end{figure}

\subsection{Evolution of Sharp Emission Lines}

Our extensive coverage - temporal and spectroscopic - encouraged us
to examine fully the rich sharp line spectrum.
 Prior to 1995 October 18 and after 1996 May 5,
the emission was too weak for reliable measurement
and, where present, superimposed on a photospheric
line.
 Observed line profiles
were decomposed into their  2 or 3 components, assumed here to
be gaussian in form,  and equivalent width, FWHM, and velocity
were measured using IRAF routines for each component.
As far as possible, the same lines were measured on each spectrum
from 1995 October 18 to 1996 May 5.
 Decomposition of a complex
profile implies perhaps that each component is physically distinct. This is
not necessarily so as the emitting region may be a single geometrical
structure with three dominant regions in relative motion. Since the C2 and C3
and possibly the C1 components are present throughout at  fixed relative
 velocities,
the structures they represent would appear to be physically related in some
sense, e.g., cloudlets on a uniformly expanding or rotating ring.

The  dataset is applied to answering the following
questions relating to the appearance of the sharp emission lines:
\begin{itemize}
\item
By how much are the components reddened? Obviously, some reddening
is interstellar and some circumstellar in origin. Detection of
a difference in the reddening of different components or a change
during the decline would be
exciting clues to the relative location of gas and dust.

\item
What are the components' velocities? Is there a change in velocity
during the decline? 
Do the radial velocities and line widths
 vary systematically with the type of the
line, i.e., neutral atom or singly-charged ion, low or high upper state
 excitation potential? Demonstrable variations or the lack of them are
clues to the relative locations of the emitting gas and the freshly
made dust.

\item
How do the  fluxes of the components change during the decline? Is there
a reduction in flux that is correlated with the fading of R\,CrB during the
decline? Again, these measurements offer clues to the locations
of emitting gas and absorbing dust.

\item
What are the physical conditions  of the emitting regions? Estimates of
temperature and electron density are provided fairly directly from relative
fluxes of small or large sets of the emission lines. 

\end{itemize}

The answers to these questions provide clues to the location of the
emitting gas in and around R\,CrB, as we discuss in Section 9.5.

\subsubsection{Reddening}

In decline, the reduction in flux from the photosphere of a R\,CrB star is 
largest in the blue and least in the red (cf. Clayton 1996). This
observation implies not unexpectedly
 that photospheric radiation is dimmed and reddened
by small dust particles. Significantly, the emission
lines appear to be unaffected by this reddening 
(cf. Clayton 1996). In the
case of R\,CrB, Payne-Gaposchkin (1963) commented ``the chromosphere
has continued to decline in brightness but is affected slightly (if at all)
by the reddening that alters the energy distribution''.
An interpretation of this result is that an optically thick dust cloud
partially obscures the emitting region; the observed emission comes
from the unobscured (i.e, unreddened) parts of the emitting region.
 Of course, the sharp
emission lines are subject to interstellar and circumstellar reddening
but this has been shown to be small for R\,CrB: E$_{\rm{B-V}}$
$\simeq$ 0.05 mag. (Rao 1974; Asplund et al. 1997).

Here, we examine whether the different emission line components are
affected  by reddening. We exploit the
fact that our spectra provide several cases of lines 
at rather different wavelengths arising from the same upper level.
Then, there is a simple
relation between the emission line {\it fluxes} of pairs of
optically thin lines

\begin{equation}
\frac {W_{\lambda}(1) F_{c}(\lambda_1)} {W_{\lambda}(2) F_{c}(\lambda_2)}
  = \frac {A(1) \lambda_2} {A(2) \lambda_1}
\end{equation}

where {\it W}$_{\lambda}$ is the equivalent width of a line,
{\it F}$_{c}(\lambda)$ is
the observed flux in the spectrum at the wavelength of the line having
the wavelength $\lambda$, and {\it A} is the transition probability
 for spontaneous
emission in the line.

\begin{table*}
\begin{minipage}{170mm}
\caption{Observed and predicted flux ratios of Ti II emission lines}
\begin{tabular}{cccccccccccccccc} \hline
\multicolumn{1}{c}{Ratio} & \multicolumn{1}{c}{Predicted} & 
\multicolumn{14}{c}{Observed ratio}\\ \cline{3-16}
\multicolumn{1}{c}{(${{\lambda}_1/{\lambda}_2}$)}&\multicolumn{1}{c}{ratio} & 
\multicolumn{14}{c}{Date}\\
& & \multicolumn{3}{c}{95 Oct 18}&\multicolumn{2}{c}{95 Nov 2}& 
\multicolumn{3}{c}{96 Jan 5}&\multicolumn{2}{c}{96 Feb 6}&
\multicolumn{3}{c}{96 Mar 2}& \multicolumn{1}{c}{96 May 5} \\
\cline{3-16}
& & \multicolumn{1}{c}{C1}&\multicolumn{1}{c}{C2}&\multicolumn{1}{c}{C3}&
\multicolumn{1}{c}{C2}&\multicolumn{1}{c}{C3}&
\multicolumn{1}{c}{C1}&\multicolumn{1}{c}{C2}&\multicolumn{1}{c}{C3}&
\multicolumn{1}{c}{C2}&\multicolumn{1}{c}{C3}&
\multicolumn{1}{c}{C1}&\multicolumn{1}{c}{C2}&\multicolumn{1}{c}{C3}&
\multicolumn{1}{c}{C2}\\
\hline
6492/4341&0.42&...  & 0.48&... & 0.21&... &...  &0.35&... & 0.42&...  &...  &...&...  &0.37 \\
8979/4341&0.10&...&0.09&...&...&...&...&...&...&...&...&...&...&...&...\\    
5189/4533&0.28&...&...&...&0.42&0.28&0.24&0.43&0.31&...&...&0.54&0.33&0.32&...\\   
6491/4534&0.022&...&...&...&0.03&...&...&0.05&...&...&...&0.10&0.04&...&0.10\\
6491/5189&0.08&...&...&...&0.07&...&...&0.11&...&0.12&0.07&0.19&0.13&0.06&0.22\\
5129/5186&1.05&0.99&0.82&1.28&...&...&...&...&...&0.91&1.20&1.00&1.05&1.10&...\\ 
\hline
\end{tabular}
\end{minipage}
\end{table*}

A set of Ti\,{\sc ii} lines was chosen. Accurate transition probabilities
were taken from Martin, Fuhr, \& Wiese (1988).
Flux ratios of red to blue lines from the same
upper state were  estimated and are compared in Table 4 with the
predicted ratios for unreddened optically thin lines. When the
Ti\,{\sc ii} lines were resolvable into 2 or 3 components, the observed
ratio was estimated separately for each component. Inspection of Table 4
shows that the observed ratios are fairly consistent with these
predictions that assume {\it no} reddening.
 The conclusion is clear. Emission lines
are very little reddened by the soot causing the decline: the limit
E$_{\rm{B-V}} \leq 0.15$ mag may be set. Note that the photometric colours
changed very little until late in the decline: Table 2 shows (B - V) $\simeq
0.5$ from 1995 October 18 to 1996 March 2 increasing to 0.9 and 1.3
on 1996 April 9 and 1996 May 5 respectively.


\subsubsection{Radial Velocities}

Radial velocities were measured for a large number of  sharp
emission lines. When possible, velocities of components C1, C2, and
C3 were recorded along with the FWHMs. Measurements were grouped by
species (neutrals and ions) and excitation potential of the
emitting level ($\chi_u$). Results are summarized in Table 5. Measurements
for selected dates are shown as a function of $\chi_u$ in Fig. 13  for
neutral atoms and Fig. 14  for singly-charged ions.

\begin{table*} 
\centering
\begin{minipage}{170mm}
\caption{Radial velocities (km s$^{-1}$) of emission lines.} 
\begin{tabular}{rcccccccc} \hline 
&&\multicolumn{3}{c}{Sharp Lines} & &\multicolumn{3}{c}{Broad Lines}\\
\cline{3-5} \cline{7-9}
\multicolumn{1}{c}{Date} &\multicolumn{1}{c}{JD} &\multicolumn{1}{c}{C1}
&\multicolumn{1}{c}{C2} &\multicolumn{1}{c}{C3} & &\multicolumn{1}{c}{He I}
&\multicolumn{1}{c}{He I} &\multicolumn{1}{c}{[N II]}\\
&\multicolumn{1}{c}{--2440000} &&&&&\multicolumn{1}{c}{7065\AA}
&\multicolumn{1}{c}{3889\AA} &\multicolumn{1}{c}{6583\AA} \\ \hline
1995 Oct 18 &10008.56 &6.0 &20.1 &32.0 &&-8.8 & ... & ...\\
Nov 2 &10023.54 &... &19.3 &35.0 &&-7.9 & ... & ... \\
Nov 12 &10033.54 &... &... &... &&-8.7 & ... &-14.4\\
1996 ~Jan 5&10088.00 &9.0 &19.0 &33.5 &&-2.6 & -5.6 &  1.1\\
Feb 6 &10119.99 &... &17.7 &32.0 &&-4.9 &-14.4 & -3.6\\
Feb 9 &10123.01 &... &... &... && ... & ... &-14.8\\
Mar 2 &10144.95 &3.8 &18.2 &32.8 &&-7.9 &-10.3 & ...\\
Apr 9 &10182.74 &8.5 &18.0 & ... &&... & ... & ...\\
May 5 &10208.85 &6.7 &18.3 &42.2 &&... &  1.1: & ...\\
\hline
\end{tabular}
\end{minipage}
\end{table*}

The mean velocity of the central and often dominant C2 
component is 18 $\pm 2$ km s$^{-1}$ corresponding to a blueshift
of about 4 km s$^{-1}$ relative to the systemic velocity.
 When the components are well sampled, the velocity
separation of C3 (red) from C2 (central) is 15 $\pm$ 2 km s$^{-1}$
and of C1(blue) from C2(central) is 11 $\pm$ 2 km s$^{-1}$.
Thus, the separations of the outer components from the central one
are approximately equal. Perhaps more importantly, the average
velocity of the C1 and C3 components is the systematic velocity to within
the errors of measurement.
 The absolute velocities evolve only slightly, if at all,  from the
first appearance of the emission lines to late in the recovery.
The observed range in velocity of the C2 component is at most about
3 km s$^{-1}$ (Table 5) declining from 20 km s$^{-1}$ on 1995 October 18
to 18 km s$^{-1}$ during and after the deepest part of the decline.
A component's radial velocity on a given date is the same 
for neutrals and ions and is independent of $\chi_u$ and line flux.
A weak velocity gradient with $\chi_u$ may be present on
occasions.
%

\begin{figure}
\epsfxsize=8truecm
\epsffile{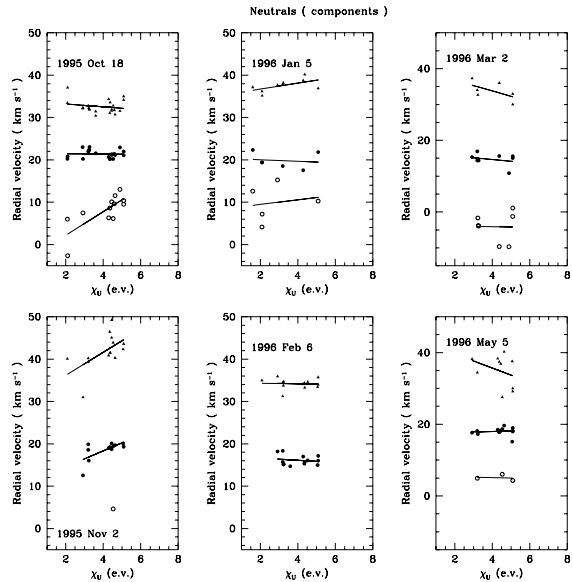}
\caption{Variation of radial velocity with upper excitation
potential ($\chi_u$) for the three components of the emission lines
of neutral atoms. Velocities of the components  C1 (blue),
 C2 (central) and C3 (red) are shown for selected dates. Straight lines
are the least-squares fits to the data.}
\end{figure}

\begin{figure}
\epsfxsize=8truecm
\epsffile{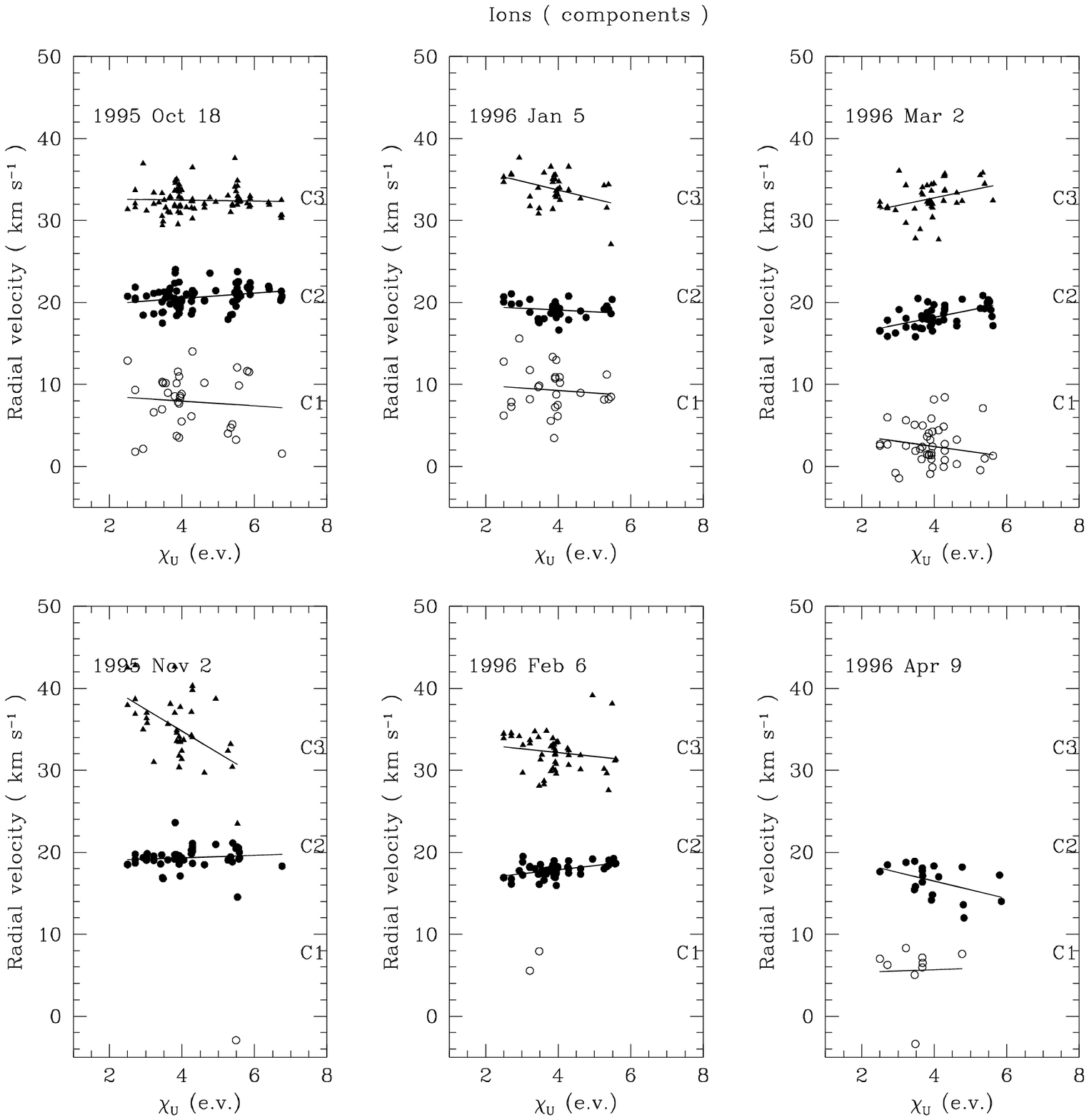}
\caption{Variation of radial velocity with upper excitation
potential ($\chi_u$) for the three components of the emission lines
of singly-charged ions. Velocities of the components  C1 (blue),
 C2 (central) and C3 (red) are shown for selected dates. Straight lines are 
the least-squares fits to the data.}
\end{figure}

\begin{figure}
\epsfxsize=8truecm
\epsffile{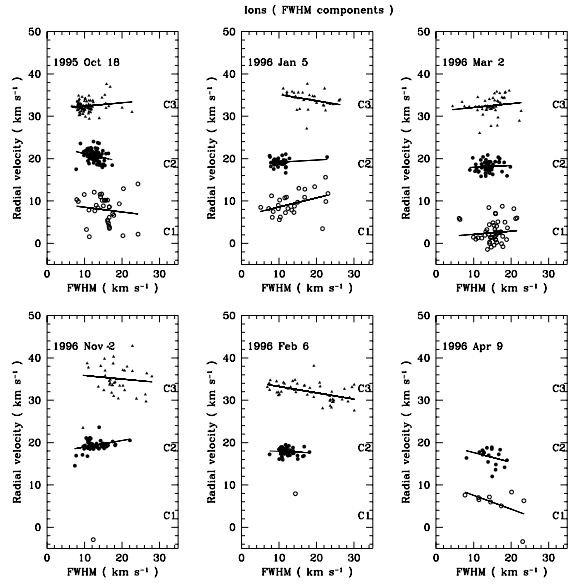}
\caption{Radial velocity versus line width (FWHM) for emission
lines of singly-charged ions for the three components C1, C2 and C3.
The straight lines are least-squares fits to the data.}
\end{figure}

Neutral and ionized lines have similar FWHM on the same spectrum.
Measurements of the ionized lines' FWHM are summarized in Fig. 15.
The few neutral lines  suitable for measurement hint at a larger
linewidth  on occasions but this may
reflect a bias in measuring broader and likely stronger neutral lines; the
difference is not more than about 3 km s$^{-1}$. 
The FWHM of C2 lines is the same for all lines independent of
species and $\chi_u$. The distribution of measurements off a single
spectrum is consistent with a very narrow distribution of the intrinsic
FWHMs.  The mean FWHM of C2 varies
very little from one observation to another and has a mean value of
10 to 14 km s$^{-1}$ uncorrected for the instrumental broadening of
nominally 5 km s$^{-1}$.
FWHM measurements for C1 and C3 show a broader distribution which 
reflects in part the difficulty of measuring these often weaker
components. On average, their mean FWHM is similar to that of the
C2 component but there are  some real differences:
C3's FWHM is less than that of C2 on 1995 October 18, and 
the distribution of FWHM for C3 and possibly C1
is distinctly broader than for C2 during the deepest part of the decline
(see Fig. 15 for 1996 January 5 and February 6).

\subsubsection{Time-dependent Flux Variations}

As Fig. 9 clearly shows a sharp line's integrated flux does not
decline in step with the diminution of the photospheric flux; a
line's equivalent width would remain constant in the event that
line and photosphere were similarly affected. The difference
between the two is striking. Measurements of selected lines
show that the integrated flux of a line is approximately the
same on 1995 October 18 as on the pre-decline maximum light
spectrum of 1995 May 18 despite the fading from V = 5.9 to  V = 10.1,
a factor of 50. By 1995 November 2 and V = 12.2, the line
flux had declined by 50\% as the star had faded a factor of 330 from maximum
light. At its faintest, the star was at V $\simeq 13.5$ or a factor of
1100 below maximum light, yet the line flux had been reduced not to 0.09\%
of its 1995 October 18 value but only to 
8\%   on 1996 January 6, 13\% on 1996 February 6, and 22\% on 1996 March 2.
This increasing trend continued, reaching 30\% on 1996 May 5. This 
extraordinary difference between `continuum' and lines is a valuable
clue to the relative locations of the obscuring soot cloud and the
emitting region of the sharp lines. (Our use of photometric magnitudes
to calibrate the continuum fluxes assumes that the lines in a photometric
bandpass do not contribute an appreciable amount of flux. This is
certainly the case for the V and R bandpasses.)

Inspection   shows that  flux ratio
of the components  C2, and C3 is not constant throughout the
decline.  At minimum light, the ratio C3/C2 is about 0.8 according to
measurements of the spectra of 1996 January 5 and February 6. Spectra
from early in the decline (1995 October 18 and November 2) and in the 
recovery phase (1996 April 9 and May 5) show a stronger C2 component
with C3/C2 of about 0.3. The spectrum of 1996 March 2 gives an intermediate
ratio of about 0.4. There is a hint in the measurements that the ratio
C3/C2 is larger for lines of low $\chi_u$.
 These  results are well shown by the
evolution of the 
Ba~II 6142\AA\ line (Fig. 10). The C1 component which is less clearly
seen may retain a constant flux ratio C1/C2 over the entire decline.

\subsection{Physical Conditions in the Emitting Region}

Ratios of line fluxes provide estimates of the physical conditions
(temperature and density) of the emitting gas. Since the line profiles are
independent of  species, the excitation
potential, and similar across the collection of spectra, it is reasonable to
suppose that the gas is reasonably homogeneous and that estimates
obtained from one species are widely applicable.

Forbidden lines are well known diagnostics. For the interpretation of the
[C\,{\sc i}] and [O\,{\sc i}] line ratios we use
a program written by Surendiranath (private communication)
 based on collision strengths
from Mendoza (1983).  Temperature T $\simeq$ 4000K and electron
density n$_e$ $\simeq 1 \times 10^7$ cm$^{-3}$ are indicated by 
the [C\,{\sc i}] ratio given in Sec. 5.2.  The corresponding [O\,{\sc i}]
ratio implies T $\simeq 5000$K and n$_e$ $\simeq 5 \times 10^7$ cm$^{-3}$.
These results are in fair agreement considering that our spectra were not
directly flux-calibrated. The emitting region is warm and fairly dense.

Fluxes of the permitted lines were analysed to obtain excitation
temperatures. We considered the species Ti\,{\sc ii}, Fe\,{\sc i},
 and Fe\,{\sc ii} that
are well represented by sharp emission lines and for which reliable
transition probabilities are available. These were taken from Martin et al.
(1988) for Ti\,{\sc ii}, and from Lambert et al. (1996)
 for Fe\,{\sc i} and Fe\,{\sc ii}.
We found that the lines were predominantly optically thin by
checking that the fluxes of lines from the same upper level were
proportional to the ratio of the lines' transition probabilities.  Lines
were assumed to be unreddened. Excitation temperatures were derived
by the procedure adopted by Pandey et al. (1996) for MV\,Sgr: 
a quantity F is derived from the line fluxes where

\begin{equation}
F = \log(W_{\lambda}F_c(\lambda)) - \log(gf\lambda^3)
\end{equation}

where $W_{\lambda}$ is an emission line's equivalent width,
and  $F_c(\lambda)$ is
the flux in the continuum at the wavelength $\lambda$ of the emission
line. 
A Boltzmann plot of logF against excitation potential of the transition's
upper level $\chi_u$ has a slope that provides the reciprocal
temperature $\theta$ = 5040/T where T is the excitation temperature.
 Sample
Boltzmann plots are shown in Fig. 16 and 17.

\begin{figure}
\epsfxsize=8truecm
\epsffile{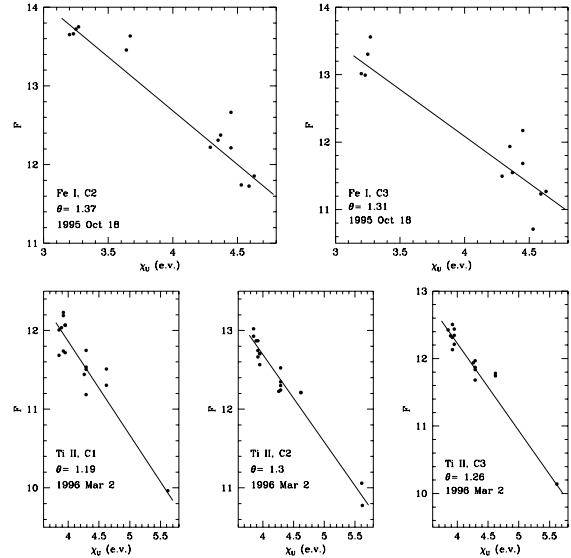}
\caption{Boltzmann plots for Fe\,{\sc i} C2 and C3 components on
1995 October 18, and Ti\,{\sc ii} C1, C2, and C3 components on
1996 March 2. }
\end{figure}

\begin{figure}
\epsfxsize=8truecm
\epsffile{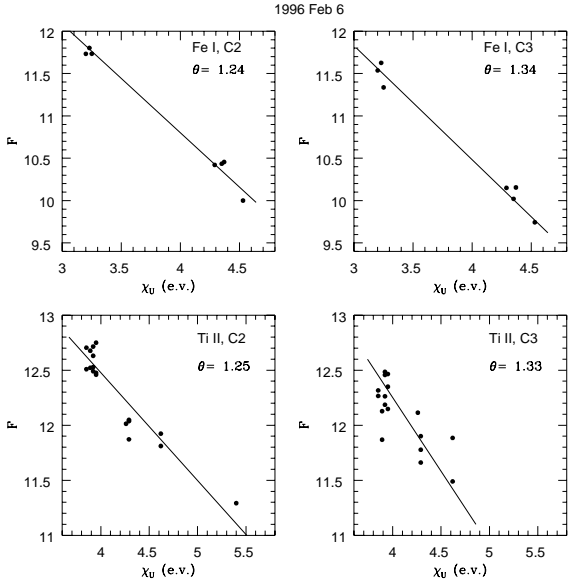}
\caption{Boltzmann plots for 1996 February 6 for the C2 and C3 components
of Fe\,{\sc i} and Ti\,{\sc ii} lines.}
\end{figure}

The striking result is that the excitation temperatures appear to be
roughly constant over the several months spanned by the observations.
Component 2, the dominant one, has an excitation temperature near 
4000K throughout the minimum. Component 3 may be systematically 
slightly cooler. Component 1 which is detectable in few
lines has a temperature close to that of C2.
The fact that lines spanning several eV
in excitation energy and from three species define a single excitation
temperature  shows that this temperature is probably close to the
kinetic temperature at a value confirmed by the [C\,{\sc i}]
line ratio and in fair agreement with that inferred from the [O\,{\sc i}]
lines.  

Knowledge of the excitation temperatures enables the ratio of neutral to
ionized iron densities to be estimated: e.g., Fe$^+$/Fe = 10$^{3.3}$
 and 10$^{3.5}$ for components C2 and C3  on
1996 February 6. There is essentially no change in these ratios
over the duration of the decline.  
Application of Saha's equation of ionization equilibrium to iron on the
assumption that ionisation and excitation are described by the same temperature
gives $n_e \simeq 3 \times 10^7$ cm$^{-3}$ to within a factor of a few. 
This range includes the $n_e$
estimate from the forbidden lines. An estimate of the total pressure
is obtainable from an application of Saha's equation to the ionization of
carbon and the assumption that C/He = 1\%: the gas pressure is
approximately 0.03 dynes cm$^{-2}$ and the gas density is about
2 $\times 10^{-9}$ g cm$^{-3}$. 
 These estimates are slight underestimates if partial
ionization of carbon is suppressed and electrons are contributed by
abundant metals. Extrapolated as an isothermal atmosphere, extension of the
photosphere corresponding to log $g$ = 0.5 (Asplund et al. 1999) reaches
the above gas pressure and density at about 10 scale heights or a mere
0.04$R_*$ above the photosphere where the gas pressure is about
500 dynes cm$^{-2}$  at Rosseland mean optical
depth $\tau_R \simeq 0.1$ (Asplund et al. 1997) and $R_*$ is the stellar
radius.  Even if carbon is fully
singly-ionized, the inferred height corresponds to only 0.05$R_*$.
Payne-Gaposchkin (1963) had earlier noted that
the temperatures of the emitting region were a little cooler than the
stellar temperature but the pressures were markedly less than
photospheric values. (Clayton et al. (1992) claim T about 6000K and $n_e$ about
$2 \times 10^{10}$ cm$^{-3}$ for V854\,Cen.
 The temperature is consistent with ours, but our above estimated
$n_e$ is substantially smaller. These estimates are derived from a set of lines
that we would expect are a mix of sharp and broad lines.
It is not
clear what T and $n_e$ mean when sharp and broad lines are mixed up.)
The observation that the sharp lines are not
fully eclipsed by the fresh soot cloud, and the size estimates from the
 [Fe\,{\sc ii}]/Fe\,{\sc ii} flux ratios and from the absolute fluxes (see
below)   show that the emitting region cannot be right above the
photosphere. Two interesting questions are obvious. Where around R\,CrB
can one expect emitting gas at a relatively high pressure? Why are the
regions immediately above the photosphere not seen in the spectrum?

An estimate of the radial distance is possible from the
ratio of forbidden and permitted Fe\,{\sc ii} lines. 
Viotti (1976) predicted the flux ratio of  4244\AA\ [Fe\,{\sc ii}] to 4233\AA\
Fe\,{\sc ii}. Essential radiative and collisional processes were included
in the statistical equilibrium calculations. Undoubtedly, better
atomic data are now available but these predictions may suffice
to indicate the approximate distance for the region of formation. Viotti
assumes that the region is irradiated by a photosphere at 10000K,
a higher temperature than is appropriate for R\,CrB. The fact that the
line fluxes and the continuum fading are nearly decoupled suggests
that  either the region is irradiated by the unobscured star throughout
the decline or the excitation of the lines is not dependent on the
receipt of photospheric radiation.

Viotti predicts the flux ratio as a function of the
dilution factor W and the customary variable n$_e$T$^{-0.5}$.
The observed ratios (see above) run from about 0.008 on 1995 October 18
to 0.13 and 0.45 on 1995 November 13 and 1996 February 6. For the appropriate
values of n$_e$T$^{-0.5}$ $\sim 10^6$, the flux ratio is
a simple function of the dilution factor and, hence, of the distance of
the region of line formation in terms of the stellar radius. The observed
flux ratios imply distance of 2, 18, and 35 stellar radii for the
three dates in question. The first estimate is probably erroneous as the
Fe\,{\sc ii} line is expected to be a blend of a sharp and a transient line. 
The other estimates indicate that the region of formation of the sharp
lines is distant from the star, a result consistent with their flux
being unaffected by the decline until a considerable fading had
occurred.

An estimate of the volume from which sharp lines are emitted may be
obtained from the emission  line fluxes. We use the [O\,{\sc i}] lines
for this purpose as their excitation is surely by electron collisions
and oxygen is expected to be neutral. If the small amount of interstellar and
circumstellar extinction is neglected, the equivalent radius of the emitting
volume is given by

\begin{equation}
\left(\frac {R_{sh}}{R_{*}}\right)^3 =
    \frac{3f_{\lambda}d^2}{n({\rm O\,{\sc i}})n_e\epsilon} \frac{1}{R_*^3}
\end{equation}

where $R_*$ is the radius of the star, $f_{\lambda}$ is the integrated
line flux, $\epsilon$ is the emission coefficient per  neutral atom and per   
electron, $d$ is the distance of the star, and $n$(O\,{\sc i}) is the density of
oxygen atoms. Adopting plausible values - $R_*$ = 100$R_{\odot}$,
$d$ = 1.35 kpc, $\epsilon$ from Surendirenath, and T $\sim 4000$ K with
$n_e \simeq 3 \times 10^7$, and an O/C abundance ratio of
10$^{-0.2}$, we find $R_{sh} \sim 1.4R_*$. This assumes that all electrons
are contributed from the ionization of carbon for which we assume
Saha's equation and local thermodynamic equilibrium.
Alternative assumptions such as full ionization of carbon atoms, or no
ionization of carbon with electrons donated by   metals do not yield
widely different estimates (say, $R_{sh} \sim (0.4 - 3) R_*$).
 It seems
clear that the region emitting the sharp lines is quite distant from the
star (say, about 20$R_*$)  and therefore very thin; if the region is a spherical
shell, its thickness is $t_{sh} \simeq 0.002R_*$ for a shell radius of
20$R_*$. This is consistent with a  single-peaked profile for the emission
lines; a thick shell would give emission lines that are
double-peaked with a separation
between the peaks of approximately twice the expansion velocity.

Comparison of the fluxes in the Ca\,{\sc ii} and [Ca\,{\sc ii}]
 lines confirms the
electron densities. At low electron densities and low radiation intensity,
the number of photons emitted in the  infrared triplet lines equal those
emitted in the two forbidden lines. On 1996 February 9, a representative
date in the deepest part of the decline, the flux in the triplet lines
is about an order of magnitude greater than that in the forbidden
lines. Since collisional de-excitation of the $^2$D state, the common
level of triplet and forbidden lines, is expected
for n$_e \sim 10^8$ cm$^{-3}$,
 the low relative intensity of the forbidden
lines is explained.
 A puzzle offered by the apparent
absence of a sharp emission line in the  H and K lines is sketched
below.

\section {Broad Emission Lines}

Discovery of broad lines may be traced back to Herbig (1949)
who found 
He\,{\sc i} 3889\AA, Ca\,{\sc ii} H and K, and the  Na\,{\sc i} D emission lines
to be broad (FWHM $\sim 170$ km s$^{-1}$).
The presence of the high excitation
 He\,{\sc i}
line in the set of broad lines implies that these lines have a
different origin to the sharp emission lines that are all
of low excitation (Payne-Gaposchkin 1963).

Our spectra greatly extend the list of broad lines.
Permitted lines 
include
the He\,{\sc i} 3889, 5876, 7065, and 10830\AA\  lines, the Na\,{\sc i} D lines,
the Ca\,{\sc ii} H and K lines and the infrared triplet lines 8498, 8542, and
8662\AA, the K\,{\sc i} lines at 7664 and 7699\AA.
 These identifications
of He\,{\sc i} lines remove all uncertainty about the correct identification
of the 3889\AA\ line (Herbig 1949; Payne-Gaposchkin 1963).
Forbidden broad lines
include the [N\,{\sc ii}] lines at 6548 and 6583\AA, the [O\,{\sc ii}] lines at
7319-7331\AA, and the [Ca\,{\sc ii}] lines at 7291 and 7323\AA.
 Our detection of 
the [O\,{\sc ii}] lines implies that Herbig's (1949, 1968) detections
of the [O\,{\sc ii}] 3727\AA\ doublet at the 1949 and 1968 deep minima refer
also to a broad line. Finally, our spectra also suggest that the C$_2$
Swan and CN Violet system bands are composed of broad lines at minimum
light.

\subsection{The He\,{\sc i}  Lines}

Neutral helium provides three lines from the triplet series: 3889\AA\
 (2s $^3$S -
2p $^3$P$^{\circ}$),
 5876\AA\ (2p $^3$P$^{\circ}$ - 3d $^3$D),
 and 7065\AA\ (2p $^3$P$^{\circ}$ -
3s $^3$S). The triplet 10830\AA\ is surely present at minimum light
but  the observed bandpass did not usually include the line.
The line was strongly in emission
on 1996 May 8 when 7065 and 5876\AA\ did not rise above the local continuum
but 3889\AA\ was obviously in emission. Other triplet lines
were searched but  not found. Singlet lines (e.g.,
 5015\AA\  2s $^1$S - 3p $^1$P$^{\circ}$),  which would be of comparable intensity
for a source in thermal equilibrium, are not present. Their absence shows,
as expected, that the He\,{\sc i} emission lines result largely from
over-population of the metastable 2s $^3$S  level.
Profiles of the three detected lines on 1996 March 2 are shown in Fig. 18.

Fig. 19 traces the evolution of the  3889\AA\ line which was seen first
(barely!) in emission on 1995 October 18 and reached its maximum equivalent
width in the 1996 January 5 spectrum.
  Spectra taken in decline prior to 1995
October 18 had not included the 3889\AA\ region so its first
appearance in emission is not known.
Emission
was present until 1996 May 8 and all signs of emission had disappeared by
1996 July 8. Unfortunately, none of the several spectra obtained
between May and July included this line.
 Emission at 3889\AA\ extends from -280 to +280 km s$^{-1}$
at the base with perhaps a slight change with line strength.
Inspection of Fig. 19 shows that sharp emission lines prominent when
the He\,{\sc i} emission line was weak do not mar the line's profile
when it was at its strongest from 1996 January to 1996 March. The reason for
this is simple and an important clue to the relative locations of the sharp
and broad  line emitting gas: the flux of  sharp emission lines decreased by
a large factor after 1995 November 2 but the broad lines' flux was
essentially unchanged throughout the decline.

The 5876\AA\ line is present in absorption at maximum light (Keenan
\& Greenstein 1963). First appearance of this line in emission occurred
on 1996 February 1; it was not present in either absorption or emission in
 1995 November 12-15.   The line was last seen in emission on 1996 March 13;
the next spectrum - 1996 April 9 - again shows neither absorption nor emission.
Since the photospheric line is a blend of He\,{\sc i}
 and C\,{\sc i} (Lambert \& 
Rao 1994), absence of an absorption implies emission filling in the 
coincident C\,{\sc i} lines
 (and adjacent lines). Absorption of about the expected
strength of the He\,{\sc i} - C\,{\sc i}
 blend was present from 1996 May 4 to the end of the
series of observations.

The 7065\AA\ line first appeared in emission above the continuum on
1995 October 18 but was  not obviously present on 1995 October 12.
 It  remained as
a distinct emission feature  until 1996 February 2; the next
spectra at 7065\AA\ from early 1996 May showed very weak emission
at the central velocity of the expected broad emission and, hence, we
suspect the broad emission at 7065\AA\ was present. Even this hint of the
broad line was absent in 1996 June and on subsequent dates.

The He\,{\sc i} emission profiles are well fitted by a Gaussian whose  width
(in km s$^{-1}$) is the same for the three lines and appears constant
over the interval in which the emission lines are prominent. For the
well observed 3889\AA\ and 7065\AA\ lines the mean widths are
322 $\pm 10$ km s$^{-1}$ and 323 $\pm 22$ km s$^{-1}$ respectively. The
radial velocities of the lines (Table 5) show that the 
emission at minimum light is systematically blue-shifted  with respect to
the mean photospheric (i.e., systemic) velocity by
 up to 31 km s$^{-1}$.
The velocity shift is constant over the period for which the broad lines
are measureable. A similar velocity shift is found for the [N\,{\sc ii}]
lines. Unfortunately, the other broad permitted and forbidden lines 
are too blended for accurate measurement of their velocity. 

 When  emission is strong (1996 February), the
7065\AA\ line is red-shifted by about 3 km s$^{-1}$ with respect to the
3889\AA\ line. This is at the limit of measurement; the shift is about
1\% of the line widths. The sense of the shift is consistent with the
supposition that the 3889\AA\ and, possibly also, the 7065\AA\ line
are optically thick: we use the wavelength of the strongest component
of these $^3$S - $^3$P triplets to compute the radial velocities. Since the
weakest component of the 3889\AA\ line is 4.2 km s$^{-1}$  to the blue
and 23 km s$^{-1}$ to the red for the 7065\AA\ line, the effect of optical
depth is to introduce a change in the lines'  effective wavelengths.

The fluxes of the He\,{\sc i} lines remained constant to within the errors of
measurement of about 30\%. This is in  contrast to the flux of
a sharp line that  declined by about a factor of ten between 1995 October 18
and 1996 January and February when the star was at its faintest.

 A striking feature of the He\,{\sc i} lines is the 
low intensity of 5876\AA\ relative to 3889\AA\ and 7065\AA.
 The  flux ratios
are measured to be
F(7065)/F(5876) $\sim 4.8$ and F(3889)/F(5876) 
$\sim  
22$.\footnote{Asplund (1995)
measured similar flux ratios at the 1993 deep minimum of RY\,Sgr:
F(7065)/F(5876) = 4.3 and F(3889)/F(5876) = 44.} We may assume that the broad
lines are little affected by reddening and certainly not by the reddening 
from dust in the cloud responsible for the decline itself.
 These ratios are quite different  from
those expected for optically thin lines produced by the
recombination of He$^+$ ions at low density: for example, prediction is that
F(7065)/F(5876) = 0.12 and F(3889)/F(5876) = 0.81 for gas at T = 10000K and
$N_e = 10^4$ cm$^{-3}$ (Aller 1984). The major discrepancy seems to be that
the 5876\AA\ flux is relatively low. Observed and predicted ratios may be
reconciled if the lines are optically thick and the electron density is high
permitting collisional excitation: Surinderinath et al.'s (1986)
calculations suggest the physical conditions T $\sim 20000$K and $n_e \sim
10^{11}$ to $10^{12}$ cm$^{-3}$. These are quite different from the
conditions of the sharp lines' emitting region.

A remarkable aspect of the He\,{\sc i} broad lines is the constancy of their
flux. Between the first appearance in mid-October 1995 to their last
appearance in early May 1996, the flux in the 3889\AA\ and in the 7065\AA\
lines was constant to within 10\%, a variation that must equal or even
better (!) the measurement errors from the rather indirect calibration
of our spectra. The constancy of flux contrasts with the case of the
sharp lines whose flux in the deepest part of the decline
(1996 January-February) fell to 10\% that measured in mid-October 1995
and out of decline.
Our flux estimates for forbidden lines, especially the [N\,{\sc ii}] 6583\AA\
line show that their flux was also constant. This condition likely applies too
to the Ca\,{\sc ii}  and Na\,{\sc i} D lines but accurate measurements are
compromised by underlying and overlying absorption. This marked difference
between the broad and sharp lines shows that their emitting regions are 
well separated. The broad line region is not occulted by the fresh cloud of
soot. 
 Equally as striking is the fact that the 3889\AA\ flux
is just 24\% less than that measured at the time of the 1962 minimum
when photographic spectra were calibrated against UBV photometry (Rao 1974).
Although additional measurements are highly desirable, it appears that the
broad line region is a long-lived feature of R\,CrB.

\begin{figure}
\epsfxsize=8truecm
\epsffile{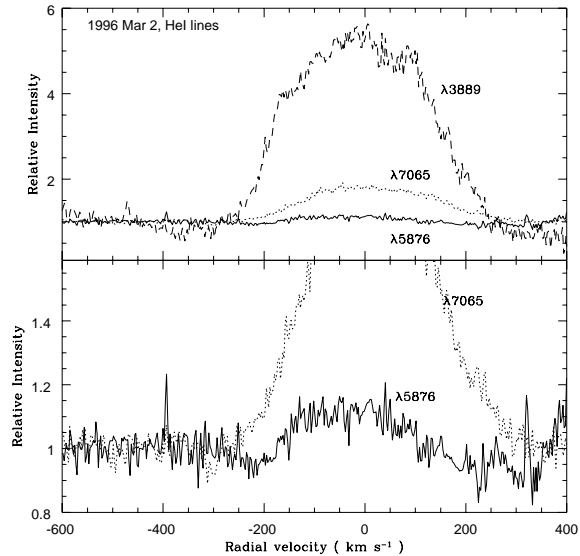}
\caption{Profiles of 3 He\,{\sc i} lines seen in the 1996 March 2 spectrum. In
the top panel, the entire profile is shown. Profiles of the 7065\AA\ and
5876\AA\ lines are shown on an expanded scale in the bottom panel.}
\end{figure}

\begin{figure}
\epsfxsize=8truecm
\epsffile{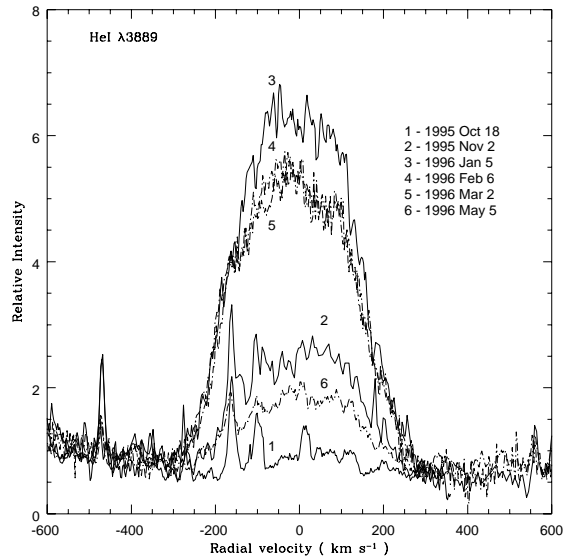}
\caption{The 3889\AA\ He\,{\sc i} profile on six occasions - see key. In the
early stages of the decline (spectra 1 and 2) and the late stages (spectrum 6)
sharp emission lines of Fe\,{\sc i} at 3886.3, 3887.0, and 3888.5\AA\ are also
visible.}
\end{figure}


\subsection{ The Permitted and Forbidden Ca\,{\sc ii} Lines}

The Ca$^+$ ion contributes 
 the 3968\AA\ and 3934\AA\ resonance (H and K) lines, the
infrared triplet of
 lines at 8542\AA, 8662\AA, and 8498\AA, and the forbidden lines at
 7291.47\AA\ and 7323.89\AA\ (Risberg 1968). 
The H and K resonance lines are a transition from the 4s $^2$S ground
state to the second excited state (4p $^2$P$^{\circ}$), which serves as the 
 upper state of the
infrared lines. The lower state of the
infrared lines, 3d $^2$D, and the ground state are connected
 by the forbidden lines. In optically thin environments,
 the ratio of the emission
coefficients of the H and K lines to the infrared triplet lines is set by a
branching ratio favouring the H and K lines. In a low density gas optically
thin to the infrared triplet lines, the number of photons emitted in the
infrared triplet lines equals the number emitted in the forbidden lines.
In light of these expectations, it is of interest to examine the
similarities and diffferences between the permitted and forbidden Ca\,{\sc ii} 
profiles and fluxes.

Broad emission in the H and K lines was seen first on 1995 October 15.
On the previous spectrum (1995 October 2) to cover these lines,
 the lines have essentially their
photospheric absorption profile. On the next spectrum - 1995 November 2 -
the lines have the basic profile that they retained throughout the
decline (Fig. 20): a broad profile with a red peak stronger than the
blue peak.
 Emission was last seen  on 1996 May 6 and the photospheric
absorption spectrum  had returned by 1996 July 8.

Emission in the infrared triplet lines was first seen on 1995 October 13
 as  {\it sharp} emission in the 8542\AA\ line; the other two lines
of the triplet were not on the  observed portion of the spectrum.
The previous spectrum - 1995 October 2 - showed
a photospheric absorption line at 8542\AA.
 The sharp emission in the triplet lines
was seen until 1996 July 26 when it was superimposed on a photospheric
absorption line. A broad component underlying the sharp feature was
seen from 1996 January 5 to 1996 March 2. 

Emission in the forbidden lines was dominated by a sharp feature  but 
weak underlying broad emission was seen on 1996 January 5 and February 6.
It is likely that this broad emission was present for a longer time but
several spectra included not the 7291\AA\ line but that at 7323\AA\ whose
broad emission, if present, is
masked  by stronger  broad  emission from  [O\,{\sc ii}] lines.
The sharp emission line in the [Ca\,{\sc ii}] lines was seen first
on 1995 October 18 and last seen on 1996 March 2. Emission was
not present on or before 1995 October 2 and on or after 1996 May 8.
The signature of emission on 1995 October 18 is clear but the spectrum of
1995 November 2 shows no signs on emission.

The flux in the broad infrared triplet lines relative to that in the H and K
lines  is approximately equal to the branching ratio and, hence, the
gas emitting the broad lines appears to be optically thin to the 
H and K lines. The flux in the forbidden lines is about a factor of 10
less than that in the infrared triplet lines which suggests that
collisional deexcitation of the $^2$D level is occurring which is
expected considering the high densities derived from the He\,{\sc i} lines. 

The H and K profiles  differ from those
of He\,{\sc i} and [N\,{\sc ii}]. While the width of
 the Ca\,{\sc ii} emission at its base is
essentially identical to that of the He\,{\sc i} and other broad lines,
the H and K lines are distinguished by a marked asymmetry: the
 red wing is a factor of two stronger than
the blue wing. This asymmetry is not due to sharp Ca\,{\sc ii}
 emission augmenting
the red wing.
 Although interstellar absorption is
present at -20 km s$^{-1}$, it is too narrow to affect the asymmetry of the
broad line. We suppose that the asymmetry is caused by absorbing gas at
a velocity of -50 to -100 km s$^{-1}$ relative to R\,CrB's systemic
velocity. This gas which is not seen in the Na\,{\sc i} D profiles may be gas
ejected in earlier declines that is photoionized by the interstellar radiation
field and now containing very few neutral sodium atoms.

An intriguing puzzle is presented by the H and K profiles is the obvious
lack of a contribution from a sharp emission line. If the gas were
optically thin to these lines, the flux of the infrared triplet lines would
imply a flux in the H and K lines that is about five times the observed
flux in the broad H and K lines  but Fig. 20 shows that the sharp line
emission, if present at all, is a small fraction of the broad line
flux.   High optical depth in the H and K lines through the
sharp line emitting region may account  in part for this puzzle. 

\begin{figure}
\epsfxsize=8truecm
\epsffile{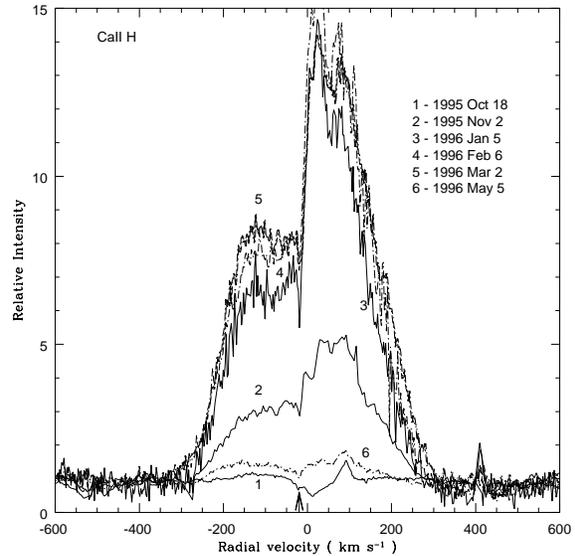}
\caption{Evolution of the Ca\,{\sc ii} H  emission line showing the asymmetric
broad line and the narrow interstellar absorption at -20 km s$^{-1}$ denoted by
the arrow.}
\end{figure}

\subsection{The K\,{\sc i} Resonance Lines}

The K\,{\sc i} resonance lines at 7664 and 7699\AA\ are present as
a composite of sharp and broad lines.\footnote{Telluric O$_2$
cross this wavelength region. Division of the R\,CrB spectrum by
that of a hot star observed at a similar airmass provides an O$_2$
-free spectrum. Owing to a small instrumental shift, the division
produces sharp inverse P Cygni profiles at the wavelengths of the
O$_2$ lines. Such a profile mars the K\,{\sc i} 7664\AA\ line.}
 The broad K\,{\sc i}
and [N\,{\sc ii}] lines have similar profiles (Fig. 21).
The velocity
shift between the broad emission lines and the
sharp lines is obvious.
The K\,{\sc i} sharp line corresponds to components C2 and C3 with no more than
a slight contribution from the  C1 component. The C2 components
of the sharp lines and probably also the broad emission lines are in the
intensity ratio of 2 to 1 for the 7664\AA\ and 7699\AA\ lines indicating
that the emitting regions are optically thin to these lines. The
C3 component appears to come from somewhat optically thick gas
because the intensity ratio is closer to 1:1 than 2:1.

\begin{figure}
\epsfxsize=8truecm
\epsffile{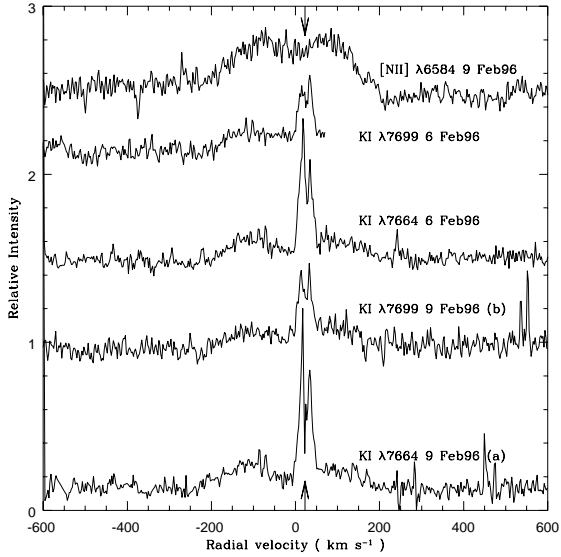}
\caption{Profiles of the K\,{\sc i} resonance lines at 7664\AA\ and 7699\AA\
from 1996 February 6. These consist of a sharp double-peaked emission line
and a broad emission line also double-peaked. The [N\,{\sc ii}] broad emission
line is shown for comparison. The K\,{\sc i} spectra have been ratioed to the
spectrum of a hot star to remove the telluric O$_2$ lines. The arrow at
22.5 km s$^{-1}$ indicates R\,CrB's mean photospheric velocity.}
\end{figure}

\subsection{The Na\,{\sc i} D Lines}

The Na\,{\sc i} D lines are the strongest features in the visual region of
R\,CrB's spectrum in decline.  Fig. 22 shows representative
profiles that are a composite of sharp and broad emission lines with sharp
interstellar absorption, and, in the recovery from minimum light, a high
velocity (blue-shifted) absorption line. The K\,{\sc i}
 and Na\,{\sc i} profiles are
quite similar; the higher abundance of Na is presumably responsible for
the appearance of absorption components in the Na\,{\sc i} D but not the K\,{\sc i}
lines. A problem peculiar to Na\,{\sc i} D is that the smaller separation between the
two resonance lines (6\AA\ versus 35\AA\ for K\,{\sc i}) results in a blending
of the broad D$_1$ and D$_2$ lines. Despite the blending, it is clear that
the velocity width of the Na\,{\sc i} D lines is quite similar to that of all other
broad lines. The sharp Na\,{\sc i} D component has contributions from C1, C2, and
C3 at all times.

\begin{figure}
\epsfxsize=8truecm
\epsffile{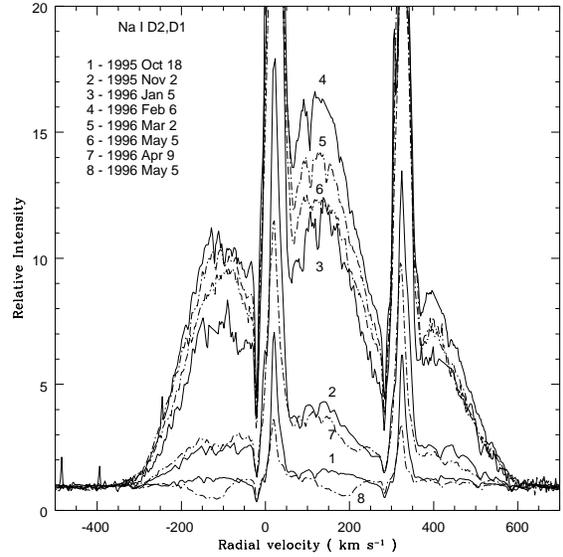}
\caption{Profiles of the Na\,{\sc i} D$_1$ and D$_2$ lines from early in the decline
to late in the recovery. All spectra are normalized to the local continuum.
The velocity scale is set for the D$_2$ line. On this scale the sharp emission
in D$_1$ appears at about 320 km s$^{-1}$. Interstellar/circumstellar 
absorption is seen just to the blue of the sharp emission line. Narrow
absorption lines at other velocities are telluric H$_2$O lines.}
\end{figure}

\subsection{The [N\,{\sc ii}] Lines}

Our spectra provide clear detections of the [N\,{\sc ii}] 6583\AA\ line. The
accompanying weaker line at 6548\AA\ is present with the
expected lower  intensity set by the branching ratio. The excited
forbidden line at 5754\AA\ was never detected. The 6548\AA\ line was
first seen on 1995 November 2 but was not present on 1995 October 18.
Emission was seen through to 1996 March 2 but not on 1996 April 9.
When the line was detected with good signal-to-noise, the profile
was clearly double-peaked (Fig. 23).

The mean velocity separation of the two peaks is 124 $\pm$ 19 km s$^{-1}$ with
no evidence of a change during the period of the observations. If treated
as overlapping gaussian profiles, the blue component is  found to be possibly
slightly broader than the red one (195$\pm$65 km s$^{-1}$ versus
163$\pm$37 km s$^{-1}$ for the FWHMs). The FWHM of the complete line
is 279 $\pm 35$ km s$^{-1}$ from 12 observations and does not change
during the decline. The [N~II] line is systematically blue-shifted
with respect to the photospheric velocity and also to the sharp 
lines. 

There are small but distinct differences between the [N\,{\sc ii}] and
 He\,{\sc i}
emission lines:
 the [N\,{\sc ii}] line is double-peaked but the He\,{\sc i} lines
are symmetrical, and the He\,{\sc i} lines are systematically broader by about
40 km s$^{-1}$. 
The fact that the
velocity offset from the photospheric lines is large and similar for
all of these broad lines suggests that their regions of line formation
are related. 

A rough estimate of the volume emitting the broad [N\,{\sc ii}] lines
is obtainable by using the lower limit to the flux ratio of the
6583\AA\ and 5755\AA\ lines. 
The analogous flux ratio of [O\,{\sc ii}] lines (see below) involves the
7320-7330\AA\ blend and 3727-3729\AA. Both lower limits imply $n_e \sim 10^6$
cm$^{-3}$ at T $\sim 5000$K. Surendiranath's emission coefficients with
the assumptions used for the calculation of the sharp line region imply
a radius of the broad line region $R_{bl} \sim 6R_*$. Unfortunately,  neither
T nor $n_e$ are well constrained. A higher $n_e$ is possible with a contraction
of the emitting volume scaling approximately as $n_e^{-2}$.

\begin{figure}
\epsfxsize=8truecm
\epsffile{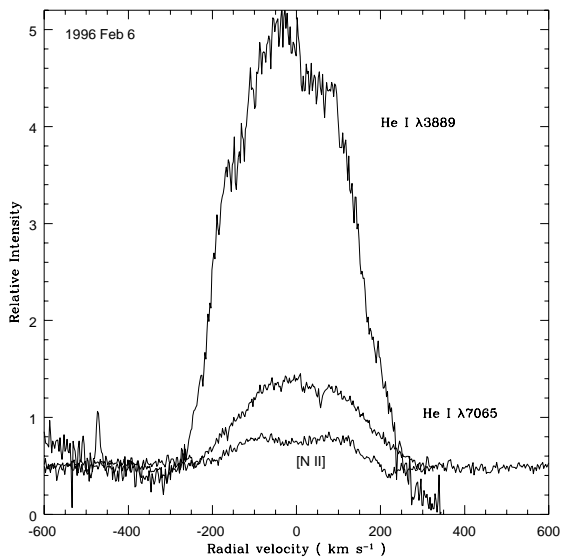}
\caption{Line profiles of [N\,{\sc ii}] 6583\AA, He\,{\sc i} 7065\AA,
and He\,{\sc i} 3889\AA\ from the spectrum of 1996 February 6.}
\end{figure}

\subsection{The [O\,{\sc ii}] Lines}

Broad emission near 7330\AA\ is primarily due to [O\,{\sc ii}] with a minor
contribution from the [Ca\,{\sc ii}] 7323\AA\ line (Fig. 11).
 The shape of the
emission indicates that it is a blend of the four expected [O\,{\sc ii}] lines:
7319.99\AA\ ($^2$D$_{5/2}$ -- $^2$P$_{3/2}$) and 7318.92\AA\ ($^2$D$_{5/2}$ --
$^2$P$_{1/2}$)  providing `one' line and 7330.73\AA\ ($^2$D$_{3/2}$ -- $^2$P$_{3/2}$) and 7329.66\AA\ ($^2$D$_{3/2}$ -- $^2$P$_{1/2}$) providing the second
`line'. (Wavelengths are taken from Moore 1993.)
These two `lines' are blended into one asymmetric broad line. The
shape of the profile suggests that the dominant contribution is made
by the pair of lines (7319.99\AA\ and 7330.73\AA) from the $^2$P$_{3/2}$
upper state, as expected. No attempt has been made to deconvolve the line
into its four [O\,{\sc ii}] and one [Ca\,{\sc ii}] components. 
The line width is quite consistent with those measured for the
[N\,{\sc ii}] 6583\AA\ line. Unfortunately, the [O\,{\sc ii}]
 3727\AA\ doublet was
never recorded with adequate signal-to-noise ratio.

\subsection{Molecular Emission Features}

{\bf C$_2$}. Absorption bands of the C$_2$ Swan system are weakly present in
the spectrum of R\,CrB at maximum light with a variable  strength.
At the onset of the decline, the C$_2$ lines go into emission:
 Fig. 24 shows  the 5165\AA\
bandhead  in emission on 1995 October 18 when transient emission lines
were strong. Emission had weakened by 1995 November 2 when the presence of
C$_2$ in absorption or emission is difficult to establish. The next
observation covering 
0-0 band shows it to be in emission (Fig. 24).
Even casual inspection of Figure 24 shows that the C$_2$ lines
 are not sharp on 1996 January 5.
 If they were, the emission spectrum would be the
inverse of the normal absorption spectrum with probably less mutilation
by atomic lines.  Striking features of the emission spectrum are that the
head is missing,  the rotational structure is unresolved, and the peak
intensity is displaced from  5165\AA, the wavelength 
 of the head in absorption, to about 5162\AA. A quite similar band profile
was present at a minimum of V854\,Cen (Rao \& Lambert 1993) and
attributed to C$_2$ line formation in cold  gas: P branch lines
of low rotational quantum number have wavelengths near 5162\AA.
The line width suggests that the regions responsible for the broad C$_2$ and
the various atomic broad lines may be related but, in view of the different
physical conditions required for C$_2$  and He\,{\sc i} emission, the regions
cannot be co-located. The C$_2$ emission consists of many blended lines
so that it is not a simple matter to derive the radial velocity of the
molecular gas. 
 The great width of the C$_2$ lines may indicate motion - turbulent
or organized - of the gas or it may arise because the emitting molecules
are between us and the scattering layer that is presumed to broaden
the photospheric lines (see below).

\begin{figure}
\epsfxsize=8truecm
\epsffile{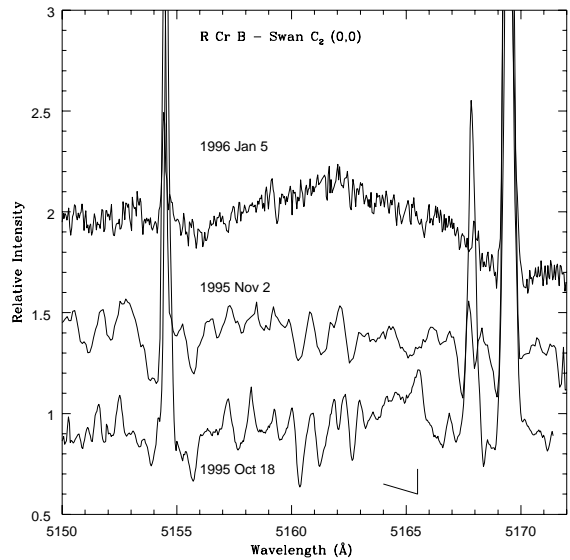}
\caption{Spectra of the C$_2$ 0-0 P branch bandhead on 1995 October 18,
1995 November 2, and 1996 January 6. Note the bandhead in emission  
on 1995 October 18, and displaced to shorter wavelengths on 1996 January 6.}
\end{figure}

A search was made for lines of the C$_2$ Phillips bands.
 The lower
electronic level for the Phillips system is the ground state of the
molecule. The Swan system's lower level is the lowest triplet level lying
slightly above the ground   (singlet) level. A combination of 
observations of Swan and Phillips bands would provide data on the
excitation of the molecule and, hence, clues to the location of the molecular
gas. Our search for Phillips lines provided nothing more than
 tantalising hints of 
absorption and emission lines. We postpone further discussion pending a full
search for 2-0 and 3-0 lines which fall in regions containing telluric
lines.

{\bf CN}. The $\Delta v$ = 0 sequence of CN violet system bands was
seen in emission first by Herbig (1949) when R\,CrB was in a deep
decline (V $\sim 14$). Additional observations were reported by
Payne-Gaposchkin (1963) and Herbig (1968). In our observations, the
$\Delta v$ = 0 sequence is first seen in emission on 1995 October 18
with a rotational structure suggesting the individual CN lines
are sharp. At the next observation of the sequence on 1996
February 6, the appearance of the band has changed dramatically:
individual lines are no longer resolved and the bandhead has shifted to the
blue.  At this time, the CN and C$_2$ bands have similar appearance.
The blue-shifting of the CN band in 
decline was noted by Herbig (1949).

\section{High-Velocity Absorption Lines}

A striking feature of the Na\,{\sc i} D profiles at earlier declines has been the
appearance of a strong broad blue-shifted absorption.
This decline was no exception: Fig. 22 shows
such absorption in the 1996 May 5 spectrum. 
Blue-shifted absorption was noted first by Payne-Gaposchkin (1963) in spectra
taken during recovery from the 1960 minimum. It was seen again in the
recovery from the 1988 minimum ( Cottrell et al. 1990;
Lambert et al. 1990).
 Clayton (1996) remarks that
high-velocity absorption lines have been seen ``early in declines and
again just before return to maximum light''. No absorption was seen
early in this decline. A similar statement applies to the 1988-1989
decline  when high-velocity absorption was seen clearly first at 100 days
after the onset and possibly at a lower velocity  at 80 days
after the onset. At these times R\,CrB was at a visual magnitude of about
10 which it had reached about 30 days into the decline.

 In this minimum,
we first detect high-velocity absorption in the 1996 April 9 spectrum when the
star was at V = 12.2 and recovering from minimum light. Comparison
of Na\,{\sc i} D profiles obtained at the same visual magnitude in the descent to
and recovery from minimum light (spectra 2 and 7 in Fig. 22) shows that the
broad emission profiles of Na\,{\sc i} D are essentially identical except for the
weak shell absorption in the spectrum taken on the recovery. As the
recovery progressed, the shell absorption changes: the equivalent
width of D$_1$ increased from 0.078\AA\ on 1996 April 9 to 0.73\AA\ on May 5 as
the mean velocity changed from -85 km s$^{-1}$ to -113 km s$^{-1}$. 
The profile is asymmetric
with the red wing steeper than the extended blue wing. The velocity of the
deepest part of the profile increases from -85 km s$^{-1}$ on 1996 April 9
to -190 km s$^{-1}$ on July 25 (Fig. 25). 
In late 1996 July, the profile split into several components . By 1996
October 2, the shell absorption had disappeared. 
  
Shell absorption is not present in the Ca\,{\sc ii} H and K lines. Superposition
of Na\,{\sc i} D and Ca\,{\sc ii} profiles  shows quite clearly that
shell absorption is absent in the latter profile. There is a hint of shell
absorption in the best K\,{\sc i} profiles.

\begin{figure}
\epsfxsize=8truecm
\epsffile{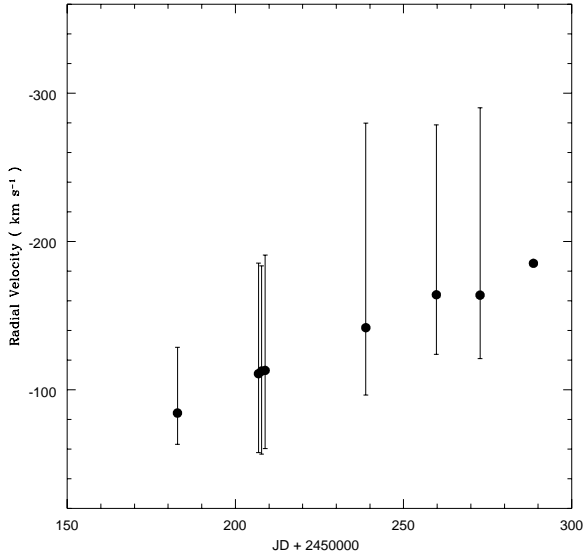}
\caption{Evolution of the Na\,{\sc i} D shell absorption velocity. The velocity
of the deepest part of the absorption is shown by the filled circles
with the vertical bars denoting the absorption's extent. For the final
point in the sequence the extent is not shown because the absorption had
broken into several distinct components.}
\end{figure}


Although reports of shell absorption are scarce, the above evolution
appears typical. Certainly, the maximum velocity in the 1988-1989 and the
present decline are similar. In the former decline, maximum velocity
 was attained in
just 100 days from onset of the decline instead of the nearly 300 days
taken in the recent decline which may reflect the fact that the
1988-1989 decline was shallower (minimum V = 11.2 against 13.6). 
Payne-Gaposchkin (1963) reported a velocity of -275 km s$^{-1}$ for shell
absorption from the 1960 minimum. This is appreciably higher than we find.
She also reported shell absorption in the Ca\,{\sc ii} H and K lines which the
recent decline did not show. The sparse dataset encourages the
view that shell absorption appears only during the recovery from minimum
light. Extrapolation of the  velocity of the maximum
absorption  and the extrema of the absorption (Figure 25) implies that
the shell's acceleration began in early 1996 January close to the mid-point
of the 150 day interval at minimum light. In the case of stars exhibiting
frequent declines (e.g., V854\,Cen), shell absorption may appear at
other phases because it may be  difficult to tie the shell absorption
to the responsible decline. Additionally, shell lines may be present
and lack an associated decline; gas from clouds off the line of sight
may intrude onto our line of sight.

\section{Photospheric Lines at Minimum Light}

Soot formation above the photosphere is considered to be the
agent responsible for the decline. As discussed above, the trigger for
soot formation appears to be seated in the photosphere. Here, we
investigate if the decline leaves a mark on the photosphere in the form
of a change of chemical composition which we determine for several dates
during the decline excluding the times when the star was at its
faintest. This exclusion is necessary because the absorption line
spectrum is then `veiled', i.e., the lines are shallow and very broad.
Discussion of the chemical composition is followed by remarks on the
veiled photospheric lines.

\subsection{Chemical Composition}

Spectra from the following four epochs were selected for measurement:
\begin{itemize}
\item
1995 September 30 - the decline began at about this time.
\item
1995 October 18 - the star had faded to V $\simeq 10$.
\item
1995 November 2 - the decline of the star's brightness had continued
and the star was close to V $\simeq 12$ at this time.
\item
1996 May 5 - recovery was underway  with the star at V $\simeq$ 10.4.
\end{itemize}

This quartet of spectra were compared with that obtained prior to the
decline on 1994 April 2 when the star was close to maximum light
in its pulsational cycle (Rao \& Lambert 1997). Comparison of the
measured equivalent widths shows that the lines in the 1995
September 30 spectrum are up to a factor of 2 stronger than in the
pre-decline maximum light spectrum. At the other three epochs, the
equivalent widths are quite similar to the pre-decline case.

Our method of analysis follows that developed for the comprehensive
abundance analysis of RCB stars (Asplund et al. 1999 - see also
Rao \& Lambert 1996). We use the
line-blanketed H-poor atmospheres computed by Asplund et al. (1997).
An abundance ratio C/He = 1\% by number is adopted.
The microturbulence is derived from Fe\,{\sc i}, and  C\,{\sc i} lines
 by the usual
condition that the derived abundance be independent of the
equivalent width. A selection of Fe\,{\sc i} and Fe\,{\sc ii} lines and the
condition of ionization equilibrium are used to fix the atmospheric
parameters T$_{\rm eff}$ and log $g$. As an additional constraint on
T$_{\rm eff}$ in 3 of the 5 cases, we use  the low excitation
 [O\,{\sc i}] 6363\AA\ line
and  high excitation permitted O\,{\sc i} lines. The parameters
T$_{\rm eff}$ = 6700 K, log $g$ = 0.0, and the microturbulence of
7 km s$^{-1}$ fit the selected indicators. That a single model
suffices is not surprising because Rao \& Lambert (1997) showed
that the atmospheric pulsation results in only small variations
of the parameters.

\begin{table*}
\centering
\begin{minipage}{140mm}
\caption{Chemical composition before and during the decline.}
 \begin{tabular}{@{}cccccccccc@{}} \hline
\multicolumn{1}{c}{Species} & \multicolumn{9}{c}{Date} \\ \cline{2-10}
& \multicolumn{1}{c}{1994 Apr 2}&\multicolumn{2}{c}{1995 Sep 30}&
\multicolumn{2}{c}{1995 Oct 18}&\multicolumn{2}{c}{1995 Nov 2}&
\multicolumn{2}{c}{1996 May 5}\\ \cline{2-10}
& \multicolumn{1}{c}{log {$\epsilon$%
\footnote{ Normalized such that
log $\Sigma$$\mu_i n_i$ = 12.15 where $\mu_i$ is the atomic weight.}}}&
\multicolumn{1}{c}{log {$\epsilon$}}&
\multicolumn{1}{c}{$\Delta$%
\footnote{$\Delta$ = [X/Fe] - {[X/Fe]}$_{1994 Apr 2}$
     where [ X/Fe] = log{($\epsilon_x/\epsilon_{Fe}$)}$_{R CrB}$ -
 log{($\epsilon_x/\epsilon_{Fe}$)}$_{\odot}$}}
&\multicolumn{1}{c}{log {$\epsilon$}}&
\multicolumn{1}{c}{$\Delta$}&\multicolumn{1}{c}{log {$\epsilon$}}&
\multicolumn{1}{c}{$\Delta$}&\multicolumn{1}{c}{log {$\epsilon$}}&
\multicolumn{1}{c}{$\Delta$}\\ \hline
 H\,{\sc i}  &  6.91   &      7.02 &-0.3   &  6.48 &-0.3  & 6.38 &-0.3  & 7.05 & 0.2\\
 Li\,{\sc i} &  2.42   &      2.93 & 0.1   & ...   & ...  & 2.33 & 0.1  & ...  & ...\\
 C\,{\sc i}  &  8.92   &      9.31 & 0.0   & ...   & ...  & 8.81 & 0.1  & 8.99 & 0.1 \\
 C\,{\sc ii} &  ...    &   ...     & ...   &  9.23 & ...  & ...  & ...  & ...  & ... \\
 N\,{\sc i}  &  8.13   &      8.40 &-0.1   &  7.66 &-0.3  & 7.91 & 0.0  & 7.80 &-0.3 \\
$[\rm {O}\,{\sc i}]$&  8.46   &      8.91 & 0.1   & ...   & ...  & ...  & ...  & 8.67 & 0.3 \\
 O\,{\sc i}  &  8.39   &      8.94 & 0.2   &  8.26 & 0.0  & 8.40 & 0.2  & 8.76 & 0.4 \\
 Na\,{\sc i} &  6.17   &      6.66 & 0.1   &  5.72 &-0.3  & 5.85 &-0.1  & 6.27 & 0.2 \\
 Mg\,{\sc i} &  6.87   &      7.38 & 0.1   & ...   & ...  & 6.81 & 0.1  & 7.10 & 0.2 \\
 Mg\,{\sc ii}&  ...    &   ...     &  ...  &  ...  & ...  & 6.49 & ...  & 6.75 & ... \\
 Al\,{\sc i} &  5.75   &      5.99 &-0.1   &  ...  & ...  & 5.65 & 0.1  & 5.70 & 0.0 \\
 Al\,{\sc ii}&  ...    &  ...      & ...   &  5.81 & ...  & ...  & ...  & 6.10 & ... \\
 Si\,{\sc i} &  7.03   &      7.08 &-0.3   &  6.47 &-0.4  & 6.69 &-0.2  & 6.85 &-0.1 \\
 Si\,{\sc ii}&  7.18   &   ...     & ...   &   ... &  ... & 6.85 & ...  & 7.14 & ... \\
 S\,{\sc i}  &  6.57   &      6.90 &-0.1   &  6.35 &-0.1  & 6.56 & 0.2  & 6.69 & 0.2 \\
 Ca\,{\sc i} &  5.23   &      5.58 & 0.0   &  ...  & ...  & 5.15 & 0.1  & 4.77 &-0.4 \\
 Fe\,{\sc i} &  6.32   &      6.71 & ...   &  ...  & ...  & 6.14 & ...  & 6.14 & ... \\
 Fe\,{\sc ii}&  6.26   &      6.63 & ...   &  6.13 & ...  & 6.06 & ...  & 6.23 & ... \\
 Ni\,{\sc i} &  5.46   &      5.82 & 0.0   &  ...  & ...  & 5.31 & 0.0  & 5.62 & 0.2 \\
 Zn\,{\sc i} &  3.81   &      4.36 & 0.2   &  3.56 &-0.1  & 3.73 & 0.1  & 4.17 & 0.4 \\
 Y\,{\sc ii} &  1.42   &      1.79 & 0.0   & ...   & ...  & ...  & ...  & ...  &...\\ 
\hline
\end{tabular}
\end{minipage}
\end{table*}

Results for the abundances are summarized in Table 6. There is a
suggestion that the abundances derived from the 1995 September 30
are consistently higher than at other times by about 0.3 to 0.5 dex.
Relative abundances X/Fe are,  however, almost identical for
this and other spectra. It is likely that the apparent difference
in composition reflects the atmospheric disturbance that subsequently 
initiated the decline. 
Since such a disturbance is not modelled
by the adopted suite of model atmospheres, the result is the
apparently anomalous abundances. Perhaps, significantly the
carbon abundance derived from the 1995 September 30 spectrum is
much closer to the adopted input abundance  than is the case
for the other spectra. In fact, the derived and input abundance
are identical to within the errors of measurement: Table 6 gives
the derived abundance as 9.3$\pm$0.3 and the input value is 9.5. At
other times, the derived carbon abundance is 0.5 to 0.7 dex less than
the input value. Since carbon dominates the continuous opacity, the
equivalent width of (weak) C~I lines should be independent of the
abundance. Asplund et al. (1999) dub the inconsistency between the
derived and input carbon abundances `the carbon problem'. After
consideration of possible resolutions of the problem, they conclude
that the key lies in the atmospheric structure. It may then be
significant that, at the time when the photosphere is disturbed, that
the carbon problem is considerably mitigated.

If the ratio X/Fe is taken as a more secure estimator of chemical
composition, it may be claimed that the photospheric composition
is unaffected by the decline. This is most probably not unexpected.
There are, however, examples of luminous supergiants with highly
peculiar compositions resulting from the efficient separation
of elements according to their condensation temperature. For example,
certain RV\,Tauri variables are  highly deficient in
Ca, Fe, and Sc but not in S and Zn (Giridhar, Lambert \& Gonzalez 1998).
Many of these stars have dusty wind and/or circumbinary diskss
 in which a separation of dust from
gas may occur. If such a separation and return of (dust-free) gas to the
atmosphere can occur for RV\,Tauri stars, might it not occur for
R\,CrB? The lack of change in the X/Fe ratios, as signified by the
quantity $\Delta$ in Table 6, suggests that it did
not occur at the 1995-1996 decline. Of course, the abundance anomalies
resulting from a dust-gas separation may differ considerably
for RV Tauri and R CrB owing to the large difference in chemical
compositions, and in pulsational amplitudes that are considerably
larger for the RV Tauri variables. In addition, the presence of high-velocity
absorption in Na\,{\sc i} D lines shows that at least some metals are in the
gas that is ejected with the dust. Sodium, however, has a relatively low
condensation temperature. Calcium has a much higher condensation temperature
so that it is intriguing that high-velocity absorption was not seen at
this decline in the Ca\,{\sc ii} H and K
lines.

\subsection{Profiles of Absorption Lines}

Previous reports of the spectra of RCBs in deep declines have noted
the weakness of the photospheric absorption lines. Adjectives
`veiled' and `diluted' have been applied to describe the appearance of
the lines. High-resolution spectra of V854\,Cen in a deep decline
showed the lines to be  absent even though the continuum
was recorded at a signal-to-noise ratio more than adequate to detect
all but the weakest lines (Rao \& Lambert 1993). Our spectra betray
the evolution of the absorption line spectrum from a normal
photospheric spectrum to a veiled spectrum at minimum light.
(In the early stages of the decline, some high-excitation lines
are filled in by emission. The apparent disappearance of these
absorption lines is not an example of veiling.)

Low excitation lines at minimum are irretrievably blended with their
own sharp emission line. Then, to study the evolution of absorption lines
we consider high excitation lines that after the disappearance of their
transient emission lines return on 1995 November 2  to their pre-decline
profile and equivalent width. On the next spectrum (1995 November 12),
these strong lines appear greatly weakened. A superior spectrum from
1996 January 5 shows the lines weakened, broadened and red-shifted (Fig. 26).
These are `veiled' lines which at this decline were present
when the star was fainter than V$ \simeq 12.5$.
  Table\,7 summarizes measurements of a set of strong
lines. 
The normal absorption lines return by 1996 May 5.

\begin{table*} 
\hoffset -0.4 true in
\centering
\begin{minipage}{170mm}
\caption{Strong absorption lines in and out of decline.}
\begin{tabular}{lccccccccccccccccc} \hline
&\multicolumn{5}{c}{Maximum Light%
\footnote{Measurements from Rao \& Lambert (1997) to show approximate range.}}& 
&\multicolumn{11}{c}{In Decline} \\
\cline{2-6} \cline{8-18}
&\multicolumn{2}{c}{1994 Apr2} &&\multicolumn{2}{c}{1994 Apr29} &
&\multicolumn{3}{c}{1995 Nov2} &&\multicolumn{3}{c}{1996 Jan5} &
&\multicolumn{3}{c}{1996 Feb6--9}\\
\cline{2-3} \cline{5-6} \cline{8-10} \cline{12-14} \cline {16-18}
\multicolumn{1}{c}{line} &\multicolumn{1}{c}{{W$_\lambda$}%
\footnote{Equivalent width in m\AA.}} &
\multicolumn{1}{c}{HW%
\footnote{Full width at half maximum line depth in km s$^{-1}$.}} &&\multicolumn{1}{c}{W$_\lambda$}&
\multicolumn{1}{c}{HW} & &\multicolumn{1}{c}{$W_\lambda$} &
\multicolumn{1}{c}{HW} &\multicolumn{1}{c}{$\Delta$V$^{\rm d}$} &
&\multicolumn{1}{c}{W$_\lambda$} &\multicolumn{1}{c}{HW} &
\multicolumn{1}{c}{$\Delta$V} & &\multicolumn{1}{c}{W$_\lambda$} &
\multicolumn{1}{c}{HW} &\multicolumn{1}{c}{$\Delta$V}\\
\hline
O\,{\sc i} 6158\AA&365&40&&343&40&&356&40&
-8&&188&...&11&&
...&
...&10\\
Si\,{\sc ii} 6347\AA&594&43&&527&43&&552&44&
-11&&265&67&15&&234&
63&13\\
Si\,{\sc ii} 6371\AA&469&34&&374&34&&471&39&
-10&&269&81&11&&139&
52&
7\\
H\,{\sc i} 6563\AA&547&52&&497&52&&425&...&
-5&&220&69&10&&
...&...&12\\
S\,{\sc i} 6743\AA&130&27&&149&27&&130&26&
-12&&...&40&...&&
...&53&13\\
S\,{\sc i} 6749\AA&184&28&&178&28&&160&27&
-11&&...&...&17&&
81&...&5\\
S\,{\sc i} 6757\AA&192&28&&196&28&&197&29&
-11&&...&40&... &&
96&75&...\\
\hline
\end{tabular}
\begin{tabular}{lccccccccccc} \hline
&\multicolumn{11}{c}{In Decline} \\
\cline{2-12}
&\multicolumn{3}{c}{1996 Mar2} &&\multicolumn{3}{c}{1996 Apr9} &
&\multicolumn{3}{c}{1996 May5}\\
\cline{2-4} \cline{6-8} \cline{10-12}
\multicolumn{1}{c}{line} &
\multicolumn{1}{c}{W$_\lambda$} &
\multicolumn{1}{c}{HW} &\multicolumn{1}{c}{$\Delta$V} &
&\multicolumn{1}{c}{W$_\lambda$} &\multicolumn{1}{c}{HW} &
\multicolumn{1}{c}{$\Delta$V} & &\multicolumn{1}{c}{W$_\lambda$} &
\multicolumn{1}{c}{HW} &\multicolumn{1}{c}{$\Delta$V}\\
\hline
O\,{\sc i} 6158\AA&248&...&11&&280&...&6&
&383&45&8\\
Si\,{\sc ii} 6347\AA&...&...&...&&392&46&
16&&...&...&... \\
Si\,{\sc ii} 6371\AA&243&64&10&&276&35&7&&479&
40&8\\
H\,{\sc i} 6563\AA& ...&...&8&&...&...&
9&&523&49&7\\
S\,{\sc i} 6743\AA &69&
36&9&&116&32&9&&157&33&
4 \\
S\,{\sc i} 6749\AA&84&37&...&&136&
31&6&&201&32&6\\
S\,{\sc i} 6757\AA &130&46&9&&...&...&...&&
224&
... &5\\
\hline
\end{tabular}
\end{minipage}
\end{table*}

\begin{figure}
\epsfxsize=8truecm
\epsffile{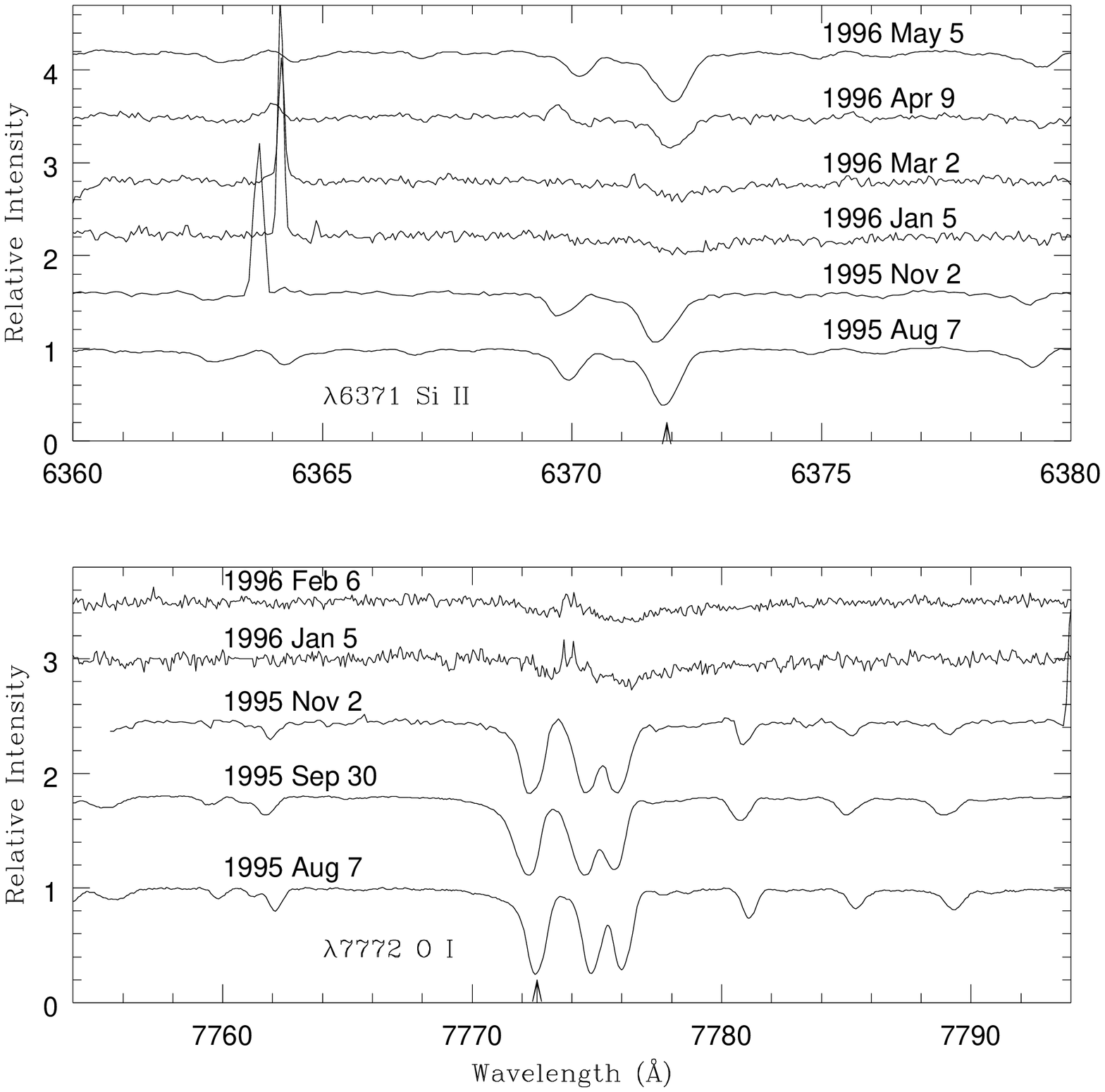}
\caption{Spectra showing the normally strong Si\,{\sc ii} 6371\AA\ line
(top panel), and the normally strong O\,{\sc i} 7774\AA\ triplet
(lower panel) on several 
occasions. All spectra are to the same  scale but are 
displaced in relative intensity for clarity.
 These lines are weakened, broadened, and redshifted on spectra
obtained from 1996 January to 1996 March.} 
\end{figure}

\section{New Insights from Spectroscopy of the 1995-96 Decline}

Spectroscopy of the 1995-1996 decline provides  detailed
evidence on R\,CrB and its environs that we now 
endeavour to relate to  the generic model 
that envisages a fresh  localized cloud
of soot forming and  obscuring the photosphere.
  Four principal spectroscopic components and/or events
should be noted. 
First, there is the photospheric disturbance occurring at the
onset, an event we refer to as the `trigger', that provided distorted
absorption line profiles and transient emission lines.
Second, there are the many sharp emission lines of neutral
and singly-ionized metals that appear largely unaffected by the decline.
Third,
there are the broad emission lines of He\,{\sc i} and other species. Fourth, 
there is a  veiling
 of the photospheric lines at minimum light.
Fifth, there is the appearance following minimum light
 of high-velocity blue-shifted
absorption in the Na\,{\sc i} D lines.
We discuss each of these  aspects.

\subsection{The  Photospheric Perturbation and Transient Emission Lines}

Spectra taken at the onset of the decline show a disturbed photosphere
with upper layers falling down and the deepest layers moving up at
a velocity greater than that expected from the extrapolation of the
previously regular pulsation. The photometric decline
appears to have occurred when
the photosphere was expected to have its maximum velocity of recession. 
In a few days following onset, the transient emission lines appear at
a velocity close to the mean photospheric velocity. We identify the
disturbed absorption lines and transient emission lines as signatures
of a shock.

Transient emission lines
appear about 2 weeks after the first observations of the
distorted photospheric lines but
incipient emission seems to be present somewhat earlier.
There must be a physical relation between the affected photospheric
absorption lines and the transient emission lines. It seems unlikely that 
the regions of formation are exactly co-located. A possibility is that
the shock to which the photospheric disturbance is attributed migrates
outward to initiate emission in lower density regions. Emission in
O\,{\sc i} high-excitation lines and in molecular C$_2$ lines is unlikely to
arise from the same layer of the atmosphere. We suppose that the
high-excitation emission lines arise from gas heated immediately behind the
shock. 
 A  possibility
suggested by theoretical  studies of pulsating R\,CrB
atmospheres is that the low atmosphere contains more than one shock;
for example, Woitke (1997) obtains steady-state solutions for a
regularly pulsating atmosphere in which a  shock  propagates outward
each pulsation but at times 2
shocks are supported by a pulsation,
and the velocity of the principal shock (relative to the
star) varies with pulsational phase
from about zero to close to the velocity $v_1$.

In models considered by Woitke (1997 -
see also Woitke, Goeres \& Sedlmayr 1996), emission lines are formed
in hot gas immediately behind a shock. Cooling times
for the hot gas appear to be very short --less than a day for a typical
model (Woitke 1997) -- and, therefore, the persistence of the transient emission
lines over a couple of weeks implies that shock propagation maintains the
emission line spectrum.
Further behind the shock in the expanded gas, densities and temperatures are
lowered with the result that cooling times are lengthened.
 It is in this post-shocked gas that conditions may be conducive
to dust formation provided the gas is sufficiently distant from the star.

 Defining the shock velocity $v_1$ to be that of the
pre-shock gas in the reference frame of the shock, Woitke's calculations
for a R\,CrB-like star show that dust may form closer to the star for
higher velocities of the outward moving shock: for example,
 $v_1$ = 50 km s$^{-1}$ provides for
dust formation as close as 0.5$R_*$ to the stellar photosphere but
at $v_1$ = 20 km s$^{-1}$ dust cannot be formed closer than about 3$R_*$
to the surface. The velocity of the emitting gas relative to the
star, $v_*$, depends on the velocity of the shock through the gas and the
velocity of  this gas relative to the star. If the atmosphere is static,
$v_* \sim v_1$. (Woitke [1997] finds a velocity of  $v_* \simeq v_1/2$ for
the case of a representative periodic shock.) The observed radial
velocity of the emission lines depends on the location and size of the shocked
region on the earth-facing hemisphere. A small region near the centre of
the hemisphere will give emission lines at the shock's velocity but lines
from an off-centre region will have a smaller radial velocity that
vanishes for a region at the limb. Emission from a large region will
have a mean velocity between $v_*$ and zero.  
If the shock propagates through infalling gas, $v_*$
 will be reduced by the infall velocity.
In the  1995 decline, there
is evidence from the photospheric line profiles at the onset
 and from the red-shifted
absorption feature accompanying the transient emission lines that
gas was infalling at about 30 km s$^{-1}$ relative to photospheric
lines.  This gas may be ahead
of the `dusty' shock or possibly related to a deeper inward moving
shock. The transient emission lines are typically redshifted by about
7 to 17 km s$^{-1}$ relative to the group A absorption lines that are
blue-shifted by about 10 km s$^{-1}$ relative to the predicted photospheric
velocity. These velocities seem to imply passage of a shock through infalling
gas.

Observations suggest a causal relation between
 the appearance of the shock 
and the onset of
the decline. The delay between the occurrence of the shock in the
photosphere and the formation of dust higher in the atmosphere appears
to have been short. However, the lack of spectra between 1995 August 9
and September 30 means that the birthdate of  the photospheric shock is
{\it not} known with precision. The presence of infalling gas prompts 
speculation that the trigger preceded the decline. Perhaps, the previous
pulsation began the sequence of events that led to the decline. 
If it can be shown that the appearance of the photospheric shock and the
onset of the decline are contemporaneous, it would seem that two shocks have
to be invoked because dust cannot form behind a photospheric
shock and that shock will require at least several days to propagate
out to the heights at which temperatures in the post-shock gas are
sufficiently cool for dust to form. In this case, one might expect to
observe different radial velocities for lines formed in the two shocks
and, hence, different radial velocities for high and low excitation
lines and evidence of two velocity components in some lines. These effects
were not seen.

R\,CrB is a regular pulsator  but declines, apparently shock initiated,
occur at random intervals. This decline occurred at maximum pulsational
velocity which seems to be the phase at which RY\,Sgr (Pugach 1977) and
V854\,Cen (Lawson et al. 1992) may also
go into decline. Evidence of photospheric shocks in the form
of line splitting has not been seen in the course of the pulsation,
but is seen in RY\,Sgr (Danziger 1963; Cottrell \& Lambert 1982;
Lawson 1986; Lawson, Cottrell \& Clark 1991; Clayton et al. 1994),
 a larger amplitude pulsator.
One supposes that a pulsation of slightly above average amplitude occurs
leading to the phenomena seen here at the 1995 onset. ( RY\,Sgr
with a larger  velocity amplitude pulsation but an otherwise
very similar atmosphere is not more prone than R\,CrB to declines, and
V854\,Cen with a small velocity amplitude is especially  susceptible to
go into decline [Lawson \& Cottrell 1997].)) 
This scenario may explain the decay of the transient emission lines. Stellar
pulsation leads to periodic development and propagation of shocks that
prove inadequate for the spawning of dust. An abnormal shock
develops, possibly related to an abnormal pulsation, and dust grows in the
unusually cool post-shock gas. This shock propagates outwards spawning dust.
Transient emission lines come from lower altitudes
primarily where gas densities are higher. At the passage of the next and
regular shock, the gas is cycled through the shock but the post-shock
gas is now warmer and unable to sustain  dust growth or the flux in the
transient emission lines. If the abnormal shock is not repeated, transient
emission lines may last no more than about a single pulsational period of
40 days, as is observed. Dust formation continues at higher altitudes as the
abnormal shock moves up, and the next normal shock encountering cooler
and dustier gas than usual may also spawn dust.

If the seat of the trigger is the photosphere and its pulsations, there
is the hope that
continuous monitoring of R\,CrB will 
establish the time delay between a photospheric
disturbance and the onset of a photometric decline. Such monitoring
would also reveal whether the 1995 decline was typical or not. If the trigger
is seated higher in the atmosphere, its first appearance may be
unobservable spectroscopically except perhaps in the ultraviolet. 

\subsection{The Broad Emission Lines: Is R CrB  a binary?}

The broad emission lines betray the presence of hot dense high-velocity
gas whose siting in the atmosphere of R\,CrB itself raises many
problems. Our alternative site is an  
accretion disk around a compact companion to the supergiant, i.e.,
we propose that R\,CrB is a spectroscopic binary. Our
interpretation was suggested in part by the discovery of broad emission
lines in the spectrum of the white dwarf  
Mira B  (Joy 1926, 1950).
Deutsch (1958) recognized that (H-rich) gas captured from the 
wind fed by  Mira A, the long-period
variable, could be responsible for the emission lines. Warner (1972),
Livio \& Warner (1984), and Reimers \& Cassatella (1985)
invoked an accretion disk around Mira B as the site of the emission
lines.   

 Introduction of the secondary and its accretion disk 
solves two principal  puzzles presented by the broad lines
 - what is the excitation
source for the He\,{\sc i} lines, and why are the lines so broad?
 The fundamental
energy source is the deep gravitational potential well provided by a
white dwarf companion; gas flowing through the accretion disk is heated
by `friction'. The width, as we show below, is simply due to the
Keplerian velocity of gas in the disk close to the white dwarf.
More fundamentally,  a realization that R CrB is a binary with a compact
companion suggests that this and some other
R CrB stars may be  the fruits of stellar evolution of a binary system.


Spectroscopic similarities between the width of R CrB's broad emission lines and
those from Mira B and other systems with accretion disks around white
dwarfs provide suggestive but not conclusive proof that R CrB is also
accompanied by a disk and a white dwarf. The clearest proof would come
from radial velocity variations of R CrB itself and of the broad line
emitting region.

The broad lines are detectable only when R CrB is
about 5 or more magnitudes below maximum light. In the 1995-1996 decline
the He\,{\sc i} lines were detectable with profiles adequately defined for
a radial velocity measurement for about 6 months (Table 5).
To within the measurement errors, the velocity of the
broad lines was constant. The only other measurement of the
He\,{\sc i} lines appears to be that by Herbig (1949) of the 3889\AA\ line
in 1949 February at a  velocity 
essentially identical to  that reported here which implies that either
the observations sampled the radial velocity curve at a similar velocity or
that R\,CrB is a single star and the He\,{\sc i} lines are
emitted by gas associated with R CrB itself.

In our search for orbital motion of R CrB, we examined the 149 measurements
of radial velocity referred to in Sec. 3.1. As noted there, the complete
set suggests a pulsational period of 42.6968 days or, if the oldest 
measurements are omitted, a period of 42.7588 days. In either case, there
is little evidence for a longer-term (i.e., orbital) velocity variation.
If R CrB is a spectroscopic binary, the lack of a detectable
orbital velocity variation implies that the $\gamma$-velocity must be close 
to the mean velocity of 22 km s$^{-1}$ and the (projected)
 velocity amplitude of 
R CrB is less than about 3 km s$^{-1}$. (A notable oddity is the apparent
absence of a pulsational variation in the measurements reported by Fernie
et al. [1972]. This may be due to an unfortunate distribution of the
measurements with respect to phase. These measurements may be fitted
by the mean curve if it is displaced about 4 km s$^{-1}$ to higher
velocities. Perhaps, this displacement is the result of orbital
motion.)

If the He\,{\sc i} lines are associated
with gas moving with the velocity of the secondary, the velocity amplitude
of the secondary star is at least 30 km s$^{-1}$, the velocity difference
between the He\,{\sc i} lines and the systemic velocity in the 1995-1996
decline. These estimates of the velocity amplitudes imply a mass ratio
$M_{\rm R}/M_{\rm Sec} \geq 10$ independent of the inclination of the
orbit to the line of sight. The orbital period must be sufficiently
long that the 6-month long observations of an essentially constant radial
velocity from the He\,{\sc i} lines are consistent
with the predicted  (slow) velocity variation. Although the required period is
dependent on the orbital eccentricity, a period longer than about 2 years
would seem to be demanded.

Brightness variations at 3.5$\mu$m were reported by Stecker (1975)
to be  periodic with a period of 1100 days. This
suggestion was reconsidered by Feast et al.  (1997)
 who estimated that the period might
be 1260 days. No suggestion was made that such a period was an
orbital period. Since the warm infrared emitting dust is relatively
distant from the star,
the orbital diameter (see below) may be much smaller (say a few stellar
radii) and
`interference' between the infrared flux and
the location of the secondary seems unlikely. 
Optical polarization measurements of R\,CrB
 (Stanford et al. 1988; Clayton et al. 1997)
indicate a preferred direction for ejection and/or accummulation
of  dust which conceivably
could be an orbital plane;  Clayton et al. propose a bipolar
geometry.

 It is readily shown that a circular orbit corresponding to a period of (say)
1260 days
and a mass ratio of 10 corresponds to a separation of the two stars of about
630$R_{\odot}$, and  maximum radial velocity amplitudes of 
2.3 and 23 km s$^{-1}$ for a 2$M_{\odot}$ R CrB star and a 0.2$M_{\odot}$
companion respectively. The photospheric radius of R CrB is about 100$R_{\odot}$ which
is substantially smaller than the Roche lobe ($R \simeq 250 - 300 R_{\odot}$).
The secondary, if a compact object, is much smaller than its Roche lobe. 
These geometrical parameters are not greatly sensitive to the choice
of period and mass ratio.  
In the accretion disk that is
 fed presumably by  R CrB's wind, 
Keplerian velocities of 200 km s$^{-1}$ are
attained at about 0.5$R_{\odot}$ from the secondary. The observed velocities are
dependent on the angle of inclination to the line of sight.     
The broad optically thick  He\,{\sc i} lines are presumed to be emitted 
by these inner regions of the disk.  The broad forbidden lines are emitted
by a much more extended region, say $R_{bl} \sim 6R_* \simeq  600R_{\odot}$
if $n_e \sim 10^6$
cm$^{-3}$. In order for the broad forbidden lines to reflect, as we assume, the
orbital motion of the secondary, their emitting volume must be approximately
centred on the secondary with a radius less than the separation between the
two stars. This separation is about 6$R_*$. With a slight increase in the
assumed $n_e$, $R_{bl}$ can be made less than this separation.

The binary scenario makes qualitative sense geometrically.
The least satisfactory aspect of the scenario concerns the
stellar masses inferred from the lower limit to the mass ratio, an
observational limit that is not dependent on the assumed orbital
inclination. A mass of 2$M_{\odot}$ for R\,CrB is about the maximum
conceivable given the space distribution of these stars.  
At first glance, a white dwarf mass of less than 0.2$M_{\odot}$ seems
hardly credible. A lower mass for R\,CrB, say 1$M_{\odot}$, implies
a lower mass for the white dwarf, say 0.1$M_{\odot}$. A  mass of
0.1 - 0.2$M_{\odot}$ for a white dwarf is  less than the average
mass of a white dwarf in the field or in a cataclysmic binary
(Warner 1995). In a survey of cool white dwarfs, Bergeron, Ruiz \& Leggett
(1997) obtain masses for 108 stars of which only 8 have
masses $M < 0.4M_{\odot}$, and none have masses less than 0.25$M_{\odot}$.
A similar analysis of hotter white dwarfs (Bergeron et al. 1995) also
found a scarcity of very low mass stars. Bergeron et al. (1997) note that
some of the stars with masses less than 0.4$M_{\odot}$ are double degenerate
binaries for which their analytical technique underestimates the
mass by perhaps a factor of 2. Others appear to be truly single stars
for which ``common envelope evolution is required to produce these
low-mass degenerates because the Galaxy is not old enough to have produced
them from single star evolution.''
 This inference may be a 
a clue to the evolutionary origin of R\,CrB: is it the product of
common envelope evolution?

One can, however, find empirical support for the low secondary mass from the
binary  89\,Her (Arellano Ferro 1984; Waters et al. 1993). This
well known F2 supergiant has an infrared excess  and a spectrum rich in
`sharp' emission lines, characteristics shared with R\,CrB.  Waters et al.
remark that, on the assumption that the primary has a mass of 0.6$M_{\odot}$,
the most probable mass for the secondary is 0.086$M_{\odot}$ with
an {\it a priori} 90\% probability that the secondary's mass is less than
0.15$M_{\odot}$. The orbital period of 89\,Her is 288 days. 

The He\,{\sc i} broad lines presumably arise from the inner regions of
the accretion disk. The disk's cooler outer reaches may provide the
lower excitation lines such as the Na\,{\sc i} D lines. It is, perhaps, difficult
in this scenario to understand why the widths of all broad lines are so
similar. An alternative scenario such as a hot bi-polar wind raises
the question of what powers the wind, a question that is quite
naturally answered in the context of an accretion disk around a compact
object. 

 The binary model is  testable.
 Since declines occur at random intervals,
the broad emission lines detected during these
declines should  be detected with both positive and negative velocity
displacements relative to the systemic velocity. Studies of well observed RCBs
show that declines occur at random times and thus cannot be linked to
particular orbital  phases.

With fascination, we
note that Alexander et al. (1972) report a velocity difference of about 66
km s$^{-1}$ between the broad emission line 3889\AA\ and RY Sgr's
mean velocity. This difference was found on two occasions separated by
about 200
days. This shift is of opposite sign to that we find for R CrB. 
Spite and Spite (1979) reported broad Na\,{\sc i} D lines from RY\,Sgr
 redshifted by 37 km s$^{-1}$
relative to Na\,{\sc i} D and other sharp emission lines. At RY\,Sgr's 1993
minimum, Asplund (1995) found broad and sharp Na\,{\sc i} D lines at the same
radial velocity and remarked that `this is contrary to the findings of
Spite and Spite (1979) at the 1977 minimum'. 
Interestingly, the mean pulsational velocity also
 appears to change suggesting that
RY\,Sgr may be in orbit around an unseen companion. Alexander et al.
(1972)  show the mean velocity  between -5 and -10 km s$^{-1}$ for
1969 and -10 to -12 km s$^{-1}$ for 1970 but measurements by Lawson,
Cottrell \& Clark (1991) for 1988 give the mean velocity as -21 km s$^{-1}$.
A range of at least 10 to 15 km s$^{-1}$ is indicated over a long period. 
Thus, there is  data to suggest RY\,Sgr may be a binary.

Other reports of He\,{\sc i} lines include the detection of the 7065\AA\ line
from S Aps (Goswami et al. 1997) where the line is essentially at the
same velocity as the sharp emission lines but the broad Na\,{\sc i} D lines are 
blue-shifted by about 100 km s$^{-1}$ implying that in this star
the broad He\,{\sc i} and broad Na\,{\sc i} D lines have a different
kinematical origin. The He\,{\sc i}
3889\AA\ line was detected from U Aqr
(Bond, Luck \& Newman 1979), RS Tel (Feast \& Glass 1973), and
V\,CrA (Feast 1975). 
 With the exception of S Aps,
 radial velocity
measurements  of the He\,{\sc i} were not reported for these stars.
 In light of the fact that RCB's may have radial velocity amplitudes of 
10-20 km s$^{-1}$ from pulsations (Lawson \& Cottrell 1997), it will
require dedication to search for low amplitude orbital velocity
variations.

\subsection{The Veiled Photospheric Spectrum}

When R\,CrB faded to below V $\simeq 12.5$ the character of the
absorption line spectrum changed, as noted in Sec. 8.2. 
Key changes of character are a redshift of the line core amounting
to 10-20 km s$^{-1}$,  development of line  broadening  and
extended red wings, and a 50\% decrease in equivalent width.
These changes are qualitatively attributable to multiple scattering
of the photospheric photons by circumstellar dust that is moving
out from the star.

Van Blerkom \& Van Blerkom (1978 - see also Herbig
1969, and Kwok 1976) summarize Monte-Carlo calculations
of stellar light scattered off dust in a radially expanding, spherically
symmetric, and optically thick cloud.
If the velocity of expansion is larger than the
velocity width of a stellar line, multiple scattering via the attendant
Doppler shifts broadens and redshifts the line. When the optically thick
shell is also geometrically extended, the core of the emergent line profile
is shifted by about the shell's expansion velocity with a red wing
extending to a few times the expansion velocity. Qualitatively, this
prediction matches the profiles observed around minimum light
with the velocity of expansion indicated to be about 25 km s$^{-1}$.
On the other hand, if the cloud is optically thick but geometrically 
thin, the line is only slightly red-shifted and broadened. Van Blerkom
\& Van Blerkom give an eloquent explanation for the difference between
profiles scattered by geometrically thick and thin shells.
Spherical symmetry is probably not required, but to create a redshift,
scattering clouds must be moving away from us.
Indeed, the cloud causing the decline blocks the star so that we see
light from the far side scattered off the receding dust also on the
far side. It is perhaps surprising that the apparent redshift of the
scattered photospheric lines is so small in light of the expansion velocity
suggested by the high-velocity Na\,{\sc i} D components (next section).  
 Clouds causing the
decline
must be between us and the star.
Scattering within such clouds will impose no more than a very small
Doppler shift on the scattered photons observed at Earth.

Transition of the normal photospheric profiles to the broadened shallow
profiles  occurs when the light scattered by the extended
cloud and received at earth
 is more intense than the light that penetrates the cloud
causing the decline. If the extended cloud is a semi-permanent collection
of clouds and 
photospheric light reaches this shell from all parts of the star except 
that blocked by the cloud causing the observed decline, one
expects that no decline of R\,CrB can be fainter than about
V $\simeq 13.5$. It has been {\it assumed} previously (e.g., Whitney et al. 1992
for V854\,Cen) that the light received during a long flat decline is
not direct light absorbed by the new cloud but light that is scattered
by the other clouds. Our high resolution spectra rather directly 
confirm this assumption in the case of R\,CrB - see also Rao \& Lambert
(1993) for confirmation for V854\,Cen.

Scattering of starlight off dust will not change an absorption line's
 equivalent width
but the broadened shifted lines seen 
at minimum light have a 50\% smaller equivalent width than their
counterparts at other times. This could be an apparent change due to
very severe broadening and the loss of very extended wings due to the 
limited S/N ratio of the spectra. Observations of the high-velocity
absorption in the Na\,{\sc i} D lines suggests gas and, therefore, dust clouds
can move at 200 km s$^{-1}$. Scattering off such clouds will
broaden lines by 200 to 300 km s$^{-1}$ and result in very shallow very
broad lines that will be difficult to detect except at high S/N ratio.
Thermal emission
by the ensemble of distant   dust clouds or the cloud causing the decline is
unlikely to dilute the line spectrum. 

Additional information could be gleaned from the wavelength dependence of the
broadening and dilution of the absorption lines. Unfortunately, strong
absorption
lines at wavelengths shorter than those of the line sample considered here
are low-excitation lines whose absorption profile is obscured by overlying
sharp emission. Clayton et al. (1997) finds
  from spectropolarimetry 
near minimum light that red light is stellar light passing directly
through the new dust cloud but the blue light is scattered
off an extended dusty region, a region they identify as a bipolar flow off
R\,CrB. This identification seems at odds with our scenario but Clayton
et al.'s observation refers to early 1996 April when our spectra show
the absorption line spectrum to be returning to its normal form. 
Regrettably, spectropolarimetry at minimum light has not been
reported. However, as predicted by our model,
 the spectropolarimetric observations of V854 Cen at deep minimum
do not show a change in position angle with wavelength (Whitney et al. 1992).

\subsection{High-velocity Ejection of Gas} 

Broad absorption lines blue-shifted by 100-200 km s$^{-1}$ were
seen in the  Na\,{\sc i} D lines during the recovery from minimum
light.
  Their  presence  appears to be
related to  recovery from minimum light. In the recovery from this decline,
absorption was not seen in the
Ca {\sc ii} H and K lines implying that the ejecta were primarily neutral.   

As argued in the previous section,  direct
light passing through the cloud causing the decline is a very
minor contributor to total light at magnitudes fainter than about
12.5 in the visual. Then, the gas associated with this cloud
cannot be seen in absorption at minimum light.
 As the recovery progresses, the cloud
thins and direct light transmitted through the cloud dominates the
spectrum. This thinning is seen in the return of the photospheric 
absorption line spectrum to its normal appearance. At this stage, 
sodium atoms in the associated gas are detectable in the spectrum.
Acceleration of the gas is due to the radiation pressure on
the grains and the momentum transfer by collisions between
dust and gas. The velocity - time plot for the broad absorption
lines (Fig. 25) suggests acceleration began in early 1996 January
when the star was at minimum light. The average acceleration is about
1 km s$^{-1}$ day$^{-1}$. Perhaps fortuitously this is close to the
prediction illustrated by Fadeyev (1988 - see his Figure 4). An alternative
view  is that acceleration of the gas began earlier
but was undetected because from mid 1995 November to early 1996
April the contribution of light transmitted through the cloud to the
total light was very slight. In this picture, the initial velocity of
about 80 km s$^{-1}$ was attained over a longer period extending from
the onset of decline when dust first formed to the first appearance of the
high-velocity Na\,{\sc i} D absorption lines, an interval of about 100 days. In the
initial stages, the gas at a low expansion velocity would be difficult to
detect owing to blending with the strong sharp emission line.
A series of spectra throughout a decline that led
to a minimum in which the photosphere was directly observable throughout
should betray the acceleration of the gas in the dust cloud responsible 
for the decline.

There has been an extended debate about the
initial location of dust formation. Scenarios are usually characterized as
`near' and `far' from the star.  `Near' and `far' imply formation at about 2
 and 20 stellar radii respectively. 
The acceleration is a potential clue to the location of the dust. Fadeyev's
calculation refers to gas forming at about 20 - 25 stellar radii from the
star. Approximate equality of observed and predicted accelerations
implies that the dust and gas by early 1996 was at this distance. Our
discovery of the photospheric trigger implies that dust condensed initially
much closer to the star. All other things being equal, the radiative
force on the grains scales inversely with the square of the distance
from the star. Then, the acceleration of dust formed `near' the star would
be about 400 times greater than Fadeyev's prediction.  
Although the gas would have lagged behind the dust, one expects this
acceleration of the dust to have led to an observable high velocity in
resonance lines such as the Na\,{\sc i} D and Ca\,{\sc ii} H and K lines. This was not
observed. There are several ways to reduce the initial acceleration in 
the case that dust forms near the star. Not only does the radiative force
increase with decreasing distance from the star but so does the
drag exerted by the gas because presumably the gas density in the cloud
is higher near the star. In the initial phases of the decline when dust
formation may be a continuing process, the  dust to gas ratio is low
and the drag on the grains necessarily higher than later in the decline.
Additionally, dust formation near the star may result in a lower
dust to gas ratio yet with adequate dust to initiate a decline.
After the cloud becomes very optically thick, stellar photons will be
multiply scattered and most likely absorbed such that the outer parts of
the cloud experience a much reduced radiative force.
In short, predictions
about the initial acceleration of dust and gas formed near the star
seem presently  uncertain.  

 A change in the gas to dust ratio may occur
as the cloud disperses and recovery to maximum light commences.
Before dispersal of the cloud,
sodium atoms and calcium ions may be largely condensed out onto
the grains. As dispersal ensues, the equilibrium ratio of free
atoms to atoms condensed onto grains necessarily shifts in 
favour of the former as collisions between atoms/ions and grains
decrease in frequency as the grains separate. Moreover, photons
now penetrate the cloud and ejection of atoms from grains may
proceed at an accelerated rate.
This simple picture of an very optically thick cloud at minimum
light accounts for the correlation between the
appearance of the broad lines and the onset of the recovery phase.

\subsection{Sharp Emission Lines - A Circumbinary Disk?}

A variety of observational clues point to 
the location of the gas emitting the sharp emission lines. 
There are now strong
indications that the lines are present at all times and, in sharp contrast
to the transient lines,  are not a  product of  the decline. Perhaps most
significant is the fact that the lines, except for a drop in flux which occurs
much later than the decline of the photospheric flux, are
quite unaffected by the decline; their velocity, profile, and degree of
excitation and ionization are  unchanged from  onset to recovery.
There is no evidence for interaction between the decline's fresh
dust cloud and the emitting region from which we suppose that the
late drop in flux is caused by occultation of a part of the emitting region
and not by a  quenching of the region.

Although observational material is limited, it appears that
the results gathered here are representative of earlier declines of
R\,CrB, and of declines of RY\,Sgr, and V854\,Cen. In particular, the sharp 
emission lines 
 show a blueshift of up to about 10 km s$^{-1}$
relative to the stellar systemic velocity in all well studied cases.
 This consistency and the absence of
redshifts  shows that the asymmetry
observed at Earth is quite unlikely to result from an asymmetric distribution
of gas and dust around the star; an asymmetric distribution constructed
without knowledge of the Earth's location must provide redshifts as well
as blueshifts across a sample of RCBs.

We shall assume that the emitting gas is
distributed symmetrically with respect to either R\,CrB, the putative
companion, or the binary. The gas is sited about 20$R_*$ from the star
in a thin layer.  One possibility is that the gas is
in a  wind off R\,CrB,  the companion, or the binary. The wind
may be spherically or axially symmetric. Distribution
of the permanent dust
clouds affects the observed line flux and profile. We assume the dust
is distributed in either a spherical shell or  a
torus. A non-spherical or toroidal distribution of dust is suggested by
the observational indications for a preferred plane for dust ejection.
 Since the dust shell's radius 
$R_{dust} \simeq 100R_*$ is seemingly  much larger than the binary separation (6$R_*$?),
the distinction between a shell centred on R\,CrB or its companion is
immaterial. As viewed from Earth,  the
emission lines are blue-shifted with respect to R\,CrB. Gas moving
away from us must then be either  partially  occulted by the star and/or
seen through the dust in the extended dust cloud.

An expanding spherical shell  with atoms everywhere within it moving outwards
at a constant velocity $v_{exp}$ provides a flat-topped emission profile
for optically thin lines  with a width 2$v_{exp}$ (Beals 1931) when the
atomic emission coefficient has a profile much narrower than 2$v_{exp}$,
 and in other circumstances the observed profile
 is a convolution of the flat-topped
shell profile and that of the atomic emission coefficient. In the present
case, the two profiles would seem to be of comparable width with $v_{exp} \simeq
10$ km s$^{-1}$ so that their convolution produces a quasi-Gaussian profile.
 The observed blueshift of about 4 km s$^{-1}$ implies a relative
reduction of intensity from the receding half of the emitting region.

If gas and dust are distributed in spherical shells of radii
$R_{gas} \simeq 20R_*$ and $R_{dust} \simeq 100R_*$ respectively,
  the star occults a very small
fraction of the receding gas: fraction of gas shell that is occulted
is $f = R_*^2/(4R_{sh}^2) \simeq 1/1600$ for $R_{sh} = 20R_*$. With the dust
exterior to the gas, all regions of the gas shell are seen through the
front of the dust shell and approaching and receding regions are subject to
the same extinction. In short, this geometry does not provide a blue-shifted
emission line of the magnitude observed.
 Of course, one may suppose the gas shell's receding half
to be thinner than the approaching half but then the model is in conflict with
the seemingly universal evidence for blue-shifted sharp emission lines. A 
larger blueshift would result if the emitting gas is placed outside the dust
shell (i.e., $R_{gas} > R_{dust}$) but this seems in conflict with estimates
of the shell radii. 

If the dust is confined to a torus, a line of sight grazing the lip
of the torus provides a better view of the far side of the torus than the
nearside. If the gas inside the central hole of the torus is expanding
radially, the net profile of the visible emitting gas will be
redshifted not blueshifted. An inward flow of gas does not seem likely,
and, therefore, we reject this  model. If the gas is rotating and
the gas on the farside from the observer's view is rotating towards us,
a blue-shift is expected. But, as the direction
 of rotation must differ from one RCB to another, this geometry would provide redshifts
as often as blueshifts across a sample of RCBs. It does seem, however,
that blueshifts
are universal. We reject models with gas and dust in tori.

Our third possibility considers a spherically symmetric wind and a toroidal
distribution of dust. If the inner radius of the dust torus is less than the
inner radius of those portions of the wind emitting the sharp lines, 
the effect of the dust torus is to decrease the intensity of the 
redshifted  relative to the blueshifted photons. The effect is a maximum
for a line sight perpendicular to the torus and vanishes for lines of sight
in the plane of the torus. This geometry may be the 
simplest that meets the observational conditions including the existence
of a preferred plane for dust ejection. The line of sight to R\,CrB must
obviously lie close to the preferred plane or declines would not be
seen. Unfortunately, such a direction minimizes the blueshift.
 The allowed misalignment between plane and line of sight and, hence,
the magnitude of the blueshift, depends on the thickness of the torus, the
inner shell radius of the torus, on the degree of asymmetry in the wind,
 and the allowed ejection angle of
dust near the star. Within this parameter space, it should be possible to
generate the observed blueshift. 

An outstanding question remains - how is gas distant from the star
maintained at the physical conditions of the sharp line emitting gas.
Recall that the gas pressure is representative of that within a few pressure
scale heights of the photosphere. Is it possible that the sharp
lines are emitted by compressed  gas behind a shock
that develops in the wind far from the star? Are the broad lines with their
larger blue shifts related to the wind and its shock?

\section{Concluding Remarks}

Our study of R\,CrB's 1995-1996 decline has provided a wealth of new
data on this fascinating episode. It is noteworthy that this decline
was preceded by a long period at which the star was at or near
maximum light and, therefore, the spectroscopic phenomena are
almost certainly attributable to the 1995-1996 decline and not to
the tailends of earlier declines. This situation may be sharply
contrasted with the case of V854\,Cen which declines so frequently that
spectroscopic changes resulting from one decline may overlap
those of another.

 Onset of the 1995-1995 decline is strikingly
correlated with a disturbance of the photospheric lines followed by
the short-lived appearance of emission lines including lines of high
excitation that we refer to as transient lines and seem to be the 
E1 lines according to Alexander et al.'s (1972) use of the terms
E1 and E2.  The sharp emission or E2 lines are present throughout the
decline unchanged till the star is well into the minimum ({\it m}$_V$ about 11 to 12). Their emitting
region is remote from the star and is affected to a much lesser extent than the
photosphere by the
growth of the dust cloud responsible for the 1995 decline. Our new
 data are  especially extensive for the broad permitted and forbidden
lines. An intriguing result is the velocity shift of about 30 km s$^{-1}$
 of the broad lines
relative to R\,CrB's mean radial velocity and to the sharp emission
lines. If this  shift 
is interpreted as orbital motion of the broad lines' emitting
gas relative to R\,CrB, we argue that the secondary must be a low mass
object such as a white dwarf of about 0.2$M_{\odot}$. We noted the parallel
with another enigmatic binary, 89\,Her. If the broad He\,{\sc i} lines
are attributed to the accretion disk around the white dwarf, one is
provided with a natural explanation for the heating and velocity of the
emitting gas. We noted that our proposal that R\,CrB is a previously
unsuspected binary is testable with patience.

A key task for the
future is to measure  R\,CrB's  broad lines' radial velocity at a series of
deep declines, or, at maximum light,
 from the broad lines that are expected to be present in
ultraviolet spectra obtainable with the {\it Hubble
Space Telescope}. The C\,{\sc ii} 1335\AA\ multiplet is definitely 
present at maximum light in R\,CrB (Holm et al. 1987)
Unfortunately, the {\it IUE} spectra from 1978-1984 lack the resolution
to assign the feature to the broad line category. The fact that R\,CrB's 
1335\AA\
 flux remained sensibly constant from maximum light into a decline
at  depth of about 4 magnitudes is suggestive of an association with the
broad-line region probed by our spectra. Unfortunately, the sharp lines
in the 1995-1996 decline also exhibited near-constant flux until the star
had declined by more than 5 magnitudes.

 Optical and ultraviolet
 spectroscopy of other RCBs in decline and at maximum light
is also of great importance. Optical observations
 in deep declines where broad emission
lines may appear should be most valuable.
 Accurate velocities are needed for comparison with the
systemic velocities that, in many cases, have yet to be determined to the 
desired accuracy. The C\,{\sc ii} 1335\AA\ multiplet is present in
RY\,Sgr's spectrum (Holm \& Wu 1982; Clayton et al. 1999).
 This multiplet and C\,{\sc iv} 1550\AA\
are present in V854\,Cen's spectrum at maximum light (Lawson et al. 1999);
the photospheric contribution to the spectrum was undetectable on the 
available IUE spectra.
The  1550\AA\ line is not present in RY\,Sgr's spectrum at maximum light
above the photospheric continuum as observed with HST/STIS
 (Clayton et al. 1999). None of these observations of
 ultraviolet spectra provide the
spectral resolution to distinguish broad from sharp emission lines.
We would expect the C\,{\sc iv} 1550\AA\ doublet to belong to the
category of broad lines. The C\,{\sc ii} lines at 1335 and also at
2325\AA\  (Clayton et al. 1992; Clayton et al. 1993)  might be a
mix of broad and sharp lines.\footnote{For V854\,Cen observed at minimum light,
 Clayton et al. (1993)
claimed that low-resolution IUE spectra showed  splitting or doubling at
a particular pulsational phase. Our examination of these IUE spectra in the
final archive shows the lines to single and unsplit, a conclusion also
reached by
Lawson et al. (1999).}
 In our picture,
broad lines from an accretion disk around a secondary star
 cannot be affected by shocks in the primary's photosphere.  Observational
evidence for profile variations of a broad line
correlated with  the primary's pulsation would be incompatible with
our attribution of these lines to an accretion disk. Lawson et al. (1999) and
Clayton et al. (1992) found fading of flux in the suspected broad lines in UV region (eg. Mg\,{\sc ii} etc.)
in some declines. It is possible on occasion when the ejected dust cloud grows
to large dimensions, it could occult part of broad emission line region resulting in
some reduction in emission line flux.

At issue is not simply whether R\,CrB has a companion. The presence of a
companion may be a long awaited clue to the evolutionary origins of these
stars.
The RCB family is largely homogeneous according to chemical composition.
Asplund et al. (1999 - see also Lambert \& Rao 1994, and Rao \& Lambert
1996) show that a majority of known RCB stars have a composition quite
similar to that of the eponym. A minority have strikingly different
compositions. The similarity suggests that the majority may have a
common origin. Then, if R CrB is a binary, one might  expect the
other majority members also to be binaries.
One proposal for forming RCBs invokes a binary --  the merger of a He white
dwarf with a C-O white dwarf (Sch\"{o}nberner 1986, 1996;
 Iben, Tutukov, \& Yungelson  1996).
The pair began as main sequence stars and a common envelope event
began the merger process that may be completed by the emission of
gravitational radiation. As considered to date, this process produces
an RCB as a single star.  This  model is most unlikely to
account for the  wide binary that is conjectured here for
R\,CrB; the merger process requires contact between the white dwarfs
but the separation between R\,CrB and its companion in our model is about 
10$^5$ times larger. Then, a clear demonstration that  R\,CrB is a binary
will imply that either R\,CrB was formed by mass transfer across a wide
binary or the companion is a relatively innocent bystander. The latter
possibility will be difficult to sustain if it can be shown that
binarity is common among the majority sample. 

What is clearly needed from observers is  a concerted  {\it long term} 
exploration of 
RCBs, both in decline and at maximum light.  Ultraviolet
spectroscopy should reveal broad lines even at maximum light. 
Observations of these lines repeated at intervals may reveal the
orbital velocity variations of the secondary and its accretion disk.
Spectroscopy of RCBs in deep declines should be pursued and velocity
measurements made on the broad emission lines. These  should eventually
show the orbital velocity of the secondary. Precise and frequent
observations of the RCBs velocities may reveal the orbital velocity
of the RCB primary, but this will be difficult as the pulsational
amplitude may exceed the orbital amplitude in most cases.  The ultimate
goal is an important one: understanding the evolutionary origins of
the most enigmatic of variable stars.

\section{Acknowledgements}

This paper is dedicated to Mike Marcario who monitored R\,CrB at our
request and  alerted us to the onset of the 1995
decline. Sadly, Mike died before the paper was completed.  We thank
the AAVSO, Yu. Efimov and Don Fernie for unpublished photometry,
 Peter
Cottrell for comments on a draft of the paper,
 Suzanne Hawley for
assistance at the telescope, A. V. Raveendran and R. Surendiranath for their
assistance with calculations, and
Peter Woitke for helpful conversations about shocks.  We are 
grateful to Yaron Sheffer for providing IUE spectra at short notice and to
the referee, Geoff Clayton, for several helpful comments.
This research was
supported in part by the National Science Foundation (grant AST 9618414).

\label{lastpage}

\end{document}